\documentclass[a4paper,openacc]{rsproca_new}
\pdfoutput=1

\usepackage{authblk}
\usepackage{graphicx} % Required for inserting images
\usepackage{amsmath}
\usepackage{isomath}
\usepackage{amssymb}
\usepackage{color}
\begin{document}
\title{Spectral approaches to stress relaxation in epithelial monolayers}

\author{Natasha Cowley$^{1,2}$, Christopher K. Revell$^{1,2}$, Emma Johns$^2$, Sarah Woolner$^2$, Oliver E. Jensen$^{1,2}$}
\address{$^{1}$Department of Mathematics, University of Manchester, Oxford Road, Manchester M13 9PL, UK\\
$^{2}$Manchester Cell Matrix Centre, School of Biological Sciences, University of Manchester, Oxford Road, M13 9PT, UK}

%%%% Subject entries to be placed here %%%%
\subject{discrete mechanics, mathematical biology, soft materials}

%%%% Keyword entries to be placed here %%%%

%%%% Insert corresponding author and its email address}
\corres{Oliver Jensen\\
\email{Oliver.Jensen@manchester.ac.uk}}

\begin{abstract}
We investigate the viscoelastic relaxation to equilibrium of a disordered planar epithelium described using the cell vertex model.  In its standard form, the model is formulated as coupled evolution equations for the locations of vertices of confluent polygonal cells.  Exploiting the model's gradient-flow structure, we use singular-value decomposition to project modes of deformation of vertices onto modes of deformation of cells.  We show how eigenmodes of discrete Laplacian operators (specified by constitutive assumptions related to dissipation and mechanical energy) provide a spatial basis for evolving fields, and demonstrate how the operators can incorporate approximations of conventional spatial derivatives.  We relate the spectrum of relaxation times to the eigenvalues of the Laplacians, modified by corrections that account for the fact that the cell network (and therefore the Laplacians) evolve during relaxation to an equilibrium prestressed state, providing the monolayer with geometric stiffness.  While dilational modes of the Laplacians capture rapid relaxation in some circumstances, showing diffusive dynamics, geometric stiffness is typically a dominant source of monolayer rigidity, as we illustrate for monolayers exposed to unsteady stretching deformations. 
\end{abstract}

% \rsbreak

%%%%%%%%%% Insert the texts which can accomdate on firstpage in the tag "fmtext" %%%%%

\maketitle

\enlargethispage{20pt}

\section{Introduction}

The cell vertex model \cite{nagai2001, farhadifar2007, alt2017, fletcher2017} provides a simple description of the mechanics of spatially disordered multicellular tissues, such as the epithelia that coat developing embryos, form our skin and line many internal organs \cite{guillot2013}. The vertex model can capture elastic, viscous and plastic deformations \cite{ANB2018b, tong2022}, as well as more exotic features of spatially disordered glassy systems such as a jamming/unjamming transition \cite{bi2015}.  The model is reasonably straightforward to implement computationally, at least in two dimensions (2D), and it is readily integrated with models of biological processes such as cell division, cell signalling and cell motility \cite{koride2018}.  The vertex model's parsimony (in having a small number of material parameters) supports inference \cite{brodland2014, kursawe2017, noll2020, roffay2021, ogita2022}, providing the model with a degree of predictive power.  Together, these features make the cell vertex model a popular tool in developmental biology, and in mechanobiology more generally, with simulations bringing valuable mechanical insights to the interpretation of experimental observations.  Given its wide utility and application, it is helpful to understand and potentially exploit the model's underlying mathematical structure.

It is well established that epithelial cells are sensitive to their mechanical environment \cite{guillot2013}.  For example, stretching of an epithelium stimulates cell division and reorients divisions along the axis of stretch, helping to dissipate stress and maintain epithelial homeostasis  \cite{benham2015, wyatt2015,  gudipaty2017, nestor2019, donker2022} . The responses of epithelial tissue to mechanical force are dynamic and adaptive; for example cell circularity following a uniaxial stretch can be restored close to pre-stretch levels in 90~min via actomyosin contraction \cite{nestor2019}. Moreover, the force regimes experienced by epithelial tissues \textit{in vivo} can vary significantly, from the fast changes seen upon wounding to the slower changes seen in the embryo during tissue morphogenesis.  In order to investigate questions arising from these dynamic considerations, for example to benchmark deformation rates against intrinsic timescales of the monolayer, it is helpful to calculate the spectrum of viscoelastic relaxation times of a monolayer.  These are determined not only by cell mechanical properties but also by the geometrical arrangements of cells within a monolayer.  By incorporating the spatial disorder that is intrinsic to many natural tissues, the vertex model is particularly well suited to address the way in which geometry and topology together mediate multiscale mechanical effects.

The cell vertex model rests on two primary assumptions.  The first is that in 2D a confluent cellular monolayer can be represented geometrically as a set of polygons that tile the plane.   
This abstraction allows cell shapes to be represented in terms of the locations of the vertices of the polygons.  The second assumption is that a mechanical energy can be defined in terms of geometric features of the polygons. 
The principle of virtual work allows forces to be defined at vertices as gradients of the energy, and movement of vertices down this energy gradient (at a rate defined by a suitable model of dissipation) takes the system to the nearest equilibrium, possibly allowing cells to exchange neighbours (via so-called T1 transitions) in the process.  In what follows, we adopt a variant of a simple but widely-used energy, defined in terms of the area and perimeter of each cell, that has two dimensionless parameters; we also refine a common model of friction between cells and their substrate.  This approach is undoubtedly an oversimplification of biological reality; nevertheless a proper understanding of this foundational model is useful when introducing additional processes.

Development of a rigorous link between discrete (cell-scale) and continuous (tissue-scale) descriptions of multicellular tissues remains an open challenge, motivating a variety of upscaling approaches.  Long-wave approximations assume disparate lengthscales and slow variation of properties over scales much longer than the size of cells, and have been conducted for the vertex model in one dimension (1D) \cite{fozard2010, barry2022} or in 2D via homogenizaton \cite{murisic2015} or using probabilistic arguments \cite{grossman2022}.  Mean-field models are based either on a phenomenological macroscopic constitutive model \cite{ishihara2017, duclut2021, fielding2022} or rely on an assumption of regularity in the organisation of cells \cite{moshe2018, hernandez2022, huang2022, kim2023, kupferman2020, staddon2023}.  Two studies in particular have emphasised the diffusive nature of the dynamics: coarse-graining of a 1D analogue of the vertex model revealed diffusive dynamics in a Lagrangian frame of reference \cite{fozard2010}; while a gradient flow of particles under an energy defined by areas of a tessellation was reduced to an (Eulerian) nonlinear diffusion equation \cite{natale2023}.  In the present study, in which we investigate a monolayer undergoing viscoelastic relaxation to an equilibrium, we remain in a discrete framework, while seeking points of contact with continuum descriptions. Our spectral perspective captures evolution across a wide range of lengthscales and timescales, retaining features that elude some upscaled models.

Building on studies of granular materials \cite{satake1993, degiuli2011}, we have demonstrated previously how the geometry of the cells themselves can be exploited in order to build operators of a discrete calculus \cite{ jensen2022, jensen2020}, including spatial Laplacians defined over vertices and over cell centres.  The eigenmodes of these operators provide bases for fields defined over a cell monolayer: exploiting Helmholtz--Hodge decomposition, we used this approach to represent the mechanical stress over a monolayer in terms of scalar potentials \cite{jensen2022}.  Low-order eigenmodes are sensitive to the shape of the monolayer as a whole, while high-order modes are sensitive to the shape of individual cells, providing a natural bridge between microscopic and macroscopic features.  (An alternative approach \cite{fruleux2021} uses a Laplacian that is weighted on the basis of a localised kernel function, rather than using the detailed structural information of individual cells.)  The Laplacians identified in \cite{jensen2022} provide a representation of scalar fields defined on cells and vertices and vector fields defined on cell edges and links between cell centres, and rely intrinsically on a primal network of cells and its dual triangulation.  Below, we introduce complementary spatial Laplacians $\mathcal{L}_{\mathcal{F}}$ and $\boldsymbol{\mathcal{L}}_{\mathcal{F}}$ that are suitable for discrete vector fields defined on vertices and scalar fields on faces, or vice versa, which do not require use of a dual network. 

These discrete spatial operators provide a valuable framework for multiscale analysis but are agnostic to the physical processes being modelled.  However the vertex model is a gradient flow on a manifold defined by both the choice of mechanical energy and the model for dissipation, implying a role for model-dependent operators.  In the form of the vertex model that we consider, each cell is characterised by two scalar variables (area and perimeter), geometric attributes that are energy-conjugate to a pressure and a tension.  The mapping between cell and vertex deformations is associated with the first variation of cell area and cell perimeter with respect to vertex locations.  We will use these quantities (in a framework provided by singular value decomposition (SVD)) to construct matrix operators that allow projection of the $2N_v$ degrees of freedom of the $N_v$ vertices in $\mathbb{R}^2$ onto the fewer $2N_c$ scalar variables defined over the $N_c$ cells.  Thereby we construct a scalar Laplacian $\mathcal{L}_c^{\mathsf{G}}$ and its vector equivalent $\boldsymbol{\mathcal{L}}_v^{\mathsf{G}}$ that combine physical parameters with geometric information.  These model-dependent operators are constructed such that they have embedded within them the the purely spatial Laplacians $\mathcal{L}_{\mathcal{F}}$ and $\boldsymbol{\mathcal{L}}_{\mathcal{F}}$, as well as interactions with non-standard (and more nonlocal) spatial operators.   Additional complexity arises from the fact that the network on which the operators are constructed evolves in time.  This can arise in two ways: either through T1 transitions (ignored here), or from vertex motion associated with stress relaxation.  Were the system defined on a static domain, one might expect the eigenvalues of $\mathcal{L}_c^{\mathsf{G}}$ and $\boldsymbol{\mathcal{L}}_v^{\mathsf{G}}$ to determine the spectrum of relaxation times.  This turns out typically not to be the case for monolayers in a jammed state: evolution of the prestressed network modifies relaxation times, endowing the monolayer with so-called geometric stiffness.  

Spectra constructed from the full Hessian of the vertex model were computed by Tong \hbox {\textit{et al.}} \cite{tong2023} in a study of monolayer viscoelasticity.

A monolayer with $N_v$ vertices has $2N_v$ degrees of freedom, a $2N_v\times 2N_v$ Hessian and $2N_v$ eigenvalues in its spectrum.  Of this set, approximately one quarter showed sensitivity to large variations in a parameter $\Gamma$ (measuring resistance to perimeter change relative to area change), with the remaining modes being relatively insensitive \cite{tong2023}.  We rationalize these observations by identifying a component of the spectrum that can be associated (for very large or small values of $\Gamma$) with a set of dilational modes of $\mathcal{L}_c^{\mathsf{G}}$ and $\boldsymbol{\mathcal{L}}_v^{\mathsf{G}}$; we show how such modes have essentially diffusive dynamics.  The remainder of the spectrum is strongly influenced by the monolayer's geometric stiffness.  

In the present study, we restrict attention primarily to monolayers in a jammed phase and ignore neighbour exchanges, extrusion and cell division.  We formulate the vertex model in Section~\ref{sec:evo} using a framework that distinguishes cell-based from vertex-based evolution.  We then use SVD in Section~\ref{sec:svd} to formally identify model-dependent Laplacian operators and to identify their contribution to the overall dynamics.  This is illustrated via computation in Section~\ref{sec:results}.  A gradient-flow perspective in Section~\ref{sec:disc} suggests connections with nonlocal aggregation-diffusion models.  Technical details are provided in appendices, where we use incidence matrices to perform key calculations.

\section{The vertex model}
\label{sec:evo}

\subsection{Evolution equations for vertices and cells}

In the cell vertex model, the evolution of a planar simply-connected monolayer formed of $N_c$ cells is simulated numerically by evolving the location of $N_v$ vertices $\mathsfbfit{r}^*(t^*)\equiv(\mathbf{r}_1^*,\dots,\mathbf{r}_{N_v}^*)^\top$ embedded in $\mathbb{R}^2$ at time $t^*$, giving the model $2N_v$ degrees of freedom (stars denote dimensional quantities).  The vertices define a network of confluent polygonal cells.  We assume all cells have the same material properties, and define the dimensional energy of cell $i$ as $K_A^* \mathcal{U}(A_i^*/A_0^*)+K_L^* \mathcal{U}(L_i^*/L_0^*)$, for some positive constants $K_A^*$ and $K_L^*$.  Here $A_i^*$ and $L_i^*$ are the area and perimeter of cell $i$; $A_0^*$ and $L_0^*$ are a reference area and perimeter at which the relevant component of the energy is minimal.  Accordingly, the dimensionless function $\mathcal{U}(\theta)$ is assumed to be convex, satisfying $\mathcal{U}(1)=\mathcal{U}'(1)=0$, $\mathcal{U}''(1)=1$ and $\mathcal{U}''(\theta)>0$ for all $\theta>0$.  We rescale lengths on $\sqrt{A_0^*}$ (so that $A_i^*=A_0^* A_i$, $L_i^*=\sqrt{A_0^*}L_i$ etc.) and energy on $K_A^*$.  Then the dimensionless energy of cell $i$ becomes $U_i(A_i,L_i)=\mathcal{U}(A_i)+\Gamma L_0^2 \mathcal{U}(L_i/L_0)$, where $L_0=L_0^*/\sqrt{A_0^*}$ and $\Gamma=(K_L^*/K_A^* L_0^2)$.  $\Gamma$ measures the relative energetic cost of cell perimeter change to area change.  The pressure and tension of cell $i$ are defined as $P_i\equiv \partial U_i/\partial A_i=\mathcal{U}'(A_i)$ and $T_i\equiv \partial U_i/\partial L_i=\Gamma L_0 \mathcal{U}'(L_i/L_0)$.    

Taylor-expanding $\mathcal{U}$ about $A_i=1$ and $L_i=L_0$ gives $U_i\approx \tfrac{1}{2}(A_i-1)^2+\tfrac{1}{2}\Gamma(L_i-L_0)^2$, recovering a commonly adopted expression for cell energy \cite{farhadifar2007, ANB2018a}, for which $P_i\approx A_i-1$ and $T_i\approx \Gamma(L_i-L_0)$.  These linear pressure and tension relationships, if extrapolated, permit areas and perimeters to shrink to zero under finite energy change.  To prevent this, we take $\mathcal{U}(\theta)=\theta(\log \theta-1)$, giving
\begin{equation}
P_i(A_i)=\log A_i, \quad T_i(L_i)=\Gamma L_0 \log(L_i/L_0), 
\label{eq:PT}
\end{equation}
with $P_i'(A_i)=1/A_i$ and $T_i'(L_i)=\Gamma(L_0/L_i)$.  The rigidity of an isolated cell is provided by tension $T_i>0$ (with $L_i>L_0$) balancing pressure $P_i<0$ (with $A_i<1$).  Under (\ref{eq:PT}), cell shrinkage requires greater mechanical energy than comparative cell expansion, while recovering the classical model for $A_i$ near unity and $L_i$ near $L_0$ (so that for a perfect hexagon, $P_i=0$ and $T_i=0$ for $L_0=2\times 12^{1/4}\approx 3.72$).  This formulation (\ref{eq:PT}) accommodates heterogeneity of areas and perimeters across a monolayer, allowing significant deviations from $A_i\approx 1$ and $L_i\approx L_0$.  

Cells are characterised by two pairs of scalar attributes, making it convenient to group these quantities into vectors of length $2N_c$.  We define $\mathsf{g}(t)$ to be the vector of cell pressures and tensions (so that $\mathsf{g}\equiv(P_1,\dots, P_{N_c}, T_1, \dots, T_{N_c})^\top$), $\mathsf{s}(t)$ to be the corresponding vector of cell areas and perimeters (we write $\mathsf{s}(t)$ as shorthand for $\mathsf{s}(\mathsfbfit{r}(t))$) and let $\mathsf{s}_0\equiv(1,1,\dots, 1, L_0, L_0, \dots, L_0)^\top$ denote reference areas and perimeters.  Pressures and tensions are related to areas and perimeters through the nonlinear function 
\begin{equation}
\mathsf{g}=\mathcal{G}(\mathsf{s}; \mathsf{s}_0, \Gamma)
\label{eq:G}
\end{equation}
that captures (\ref{eq:PT}).  The total mechanical energy of the monolayer $U=\sum_i U_i$ satisfies $\dot{U}=\mathsf{g}^\top \dot{\mathsf{s}}$ and $U_\mathsf{s}=\mathsf{g}$.  Here a dot denotes $\mathrm{d}/\mathrm{d}t$ and the subscript $\mathsf{s}$ denotes a Fr\'echet derivative.  Likewise, $\dot{\mathsf{g}}=\mathsf{G}\dot{\mathsf{s}}$ where $\mathsf{G}$ is a matrix satisfying
\begin{equation}
\mathsf{G}={\mathcal{G}_\mathsf{s}}\equiv
\left(
\begin{matrix}
\mathsf{A}_c^{-1} & \mathsf{0} \\
\mathsf{0} & \Gamma L_0 \mathsf{L}_{c}^{-1}
\end{matrix} \right),
     \label{eq:nonlinG}
\end{equation}
where $\mathsf{A}_c=\mathrm{diag}(A_1,\dots,A_{N_c})$ and $\mathsf{L}_{c}=\mathrm{diag}(L_1,\dots,L_{N_c})$.  To explain notation, we use $i=1,\dots,N_c$ to sum over cells and $\alpha=1,\dots,2N_c$ to sum over pairs of attributes assigned to cells; $k=1,\dots N_v$ sums over vertices.  We use a lower-case bold font to denote vectors in the physical plane $\mathbb{R}^2$, lower-case sans serif to denote $2N_c$-vectors and sans serif italic to denote vectors or tensors with 2-vector-valued or tensor-valued components.  We adopt matrix notation but use  indices for clarity where necessary.  $\top$ (or $T$) denotes a transpose with respect to coordinates $\alpha$ or $k$ (or $\mathbb{R}^2$).

Linearization around an equilibrium state (writing $\mathsf{s}=\bar{\mathsf{s}}+\hat{\mathsf{s}}$, etc.) gives $\bar{\mathsf{g}}= \mathcal{G}(\bar{\mathsf{s}};\mathsf{s}_0, \Gamma)$ with perturbations satisfying $\hat{\mathsf{g}}\approx \bar{\mathsf{G}}\hat{\mathsf{s}}$, using (\ref{eq:nonlinG}).  The total energy, when expanded in terms of variations in areas and perimeters, is then $U\approx \sum_i \bar{U}_i + \hat{\mathsf{s}}^\top \bar{\mathsf{g}} + \tfrac{1}{2} \hat{\mathsf{s}}^\top \bar{\mathsf{G}}\hat{\mathsf{s}}$.  Accordingly, we define inner products for scalar-valued attributes $\mathsf{u}$, $\mathsf{v}$ of cells, and vector-valued attributes $\mathsfbfit{u}$, $\mathsfbfit{v}$ of vertices, as
\begin{subequations}
    \begin{equation}
\langle \mathsf{u},\mathsf{v}\rangle_{\mathsf{G}} \equiv \mathsf{u}^\top {\mathsf{G}} \mathsf{v}\equiv {\textstyle{\sum_{\alpha,\alpha'}}} u_\alpha {G}_{\alpha,\alpha'}v_{\alpha},
\quad
    \langle {\mathsfbfit{u}},\mathsfbfit{v} \rangle_{\mathsf{E}}\equiv  {\mathsfbfit{u}}^\top {\mathsf{E}}\cdot \mathsfbfit{v}\equiv  {\textstyle\sum_k} {E}_k {\mathbf{u}}_k \cdot\mathbf{v}_k. 
\label{eq:inner0}
\end{equation}
Here $\mathsf{E}=\mathrm{diag}(E_1,\dots,E_{N_v})$ and
$E_k$ is an area assigned to each vertex, defined as the area of the triangle having vertices at the edge centroids neighbouring a given vertex (see Fig.~\ref{fig:hh}a below).  While $\mathsf{G}$ is the natural weighting for areas and perimeters, $\mathsf{G}^{-1}$ is the natural weighting for pressures and tensions, so that the quadratic contribution to the energy becomes
\begin{equation}
\tfrac{1}{2}\langle \hat{\mathsf{s}} , \hat{\mathsf{s}} \rangle_{\bar{\mathsf{G}}}= \tfrac{1}{2}\langle \hat{\mathsf{g}} , \hat{\mathsf{g}} \rangle_{\bar{\mathsf{G}}^{-1}}\equiv \tfrac{1}{2} \hat{\mathsf{g}}^\top \bar{\mathsf{G}}^{-1}\hat{\mathsf{g}}.
\label{eq:secondenergy}
\end{equation}
\label{eq:innprods}
\end{subequations}

Some of the complexity of the vertex model originates in the nonlinear relationship between areas, perimeters and cell vertex locations.  We capture this by defining $\mathbf{M}_{\alpha k}=
\partial s_\alpha/\partial \mathbf{r}_k$, 
or $\mathsfbfit{M}=\mathsf{s}_\mathsfbfit{r}$.  The elements of $\mathsfbfit{M}$ related to $\partial A_i/\partial \mathbf{r}_k$ can be visualised as the normal to the link between edge centroids adajecent to vertex $k$ of cell $i$; those related to $\partial L_i/\partial \mathbf{r}_k$ correspond to unit vectors acting along the edges of cell $i$ adjacent to vertex $k$, as explained in Appendix~\ref{sec:appx}.  Going to higher order, we can relate small changes in area and perimeter to vertex displacements by 
\begin{equation}
    {\delta s_\alpha}={\textstyle\sum_k} \mathbf{M}_{\alpha k}\cdot \delta\mathbf{r}_k+\tfrac{1}{2}{\textstyle{\sum_{k,k'}}} \delta \mathbf{r}_k\cdot \mathbf{M'}_{\alpha k k'}\cdot \delta\mathbf{r}_{k'}+\dots
\label{eq:x}
\end{equation}
where $\mathbf{M'}_{\alpha k k'}\equiv\partial^2 s_\alpha/\partial \mathbf{r}_k \partial \mathbf{r}_{k'}$ (see Appendix~\ref{sec:appx}), and therefore
\begin{equation}
    \dot{s}_\alpha={\textstyle\sum_k} \mathbf{M}_{\alpha k}\cdot\dot{\mathbf{r}_k}, \quad \hbox{i.e.} \quad \dot{\mathsf{s}}=\mathsfbfit{M}\cdot \dot{\mathsfbfit{r}}.
\label{eq:sdot}
\end{equation}
Here we see a suggestion of $\mathsfbfit{M}\cdot $ acting analogously to a divergence operator, mapping a discrete (vector-valued) velocity field defined over vertices to (scalar-valued) area and perimeter changes defined over cells.  This notion will be developed more formally below.  

A common model of viscous dissipation (often chosen more for computational convenience than for mechanical realism) assigns a drag to each vertex, leading to the coupled evolution equations $\dot{\mathbf{r}}_k=-\partial U/\partial \mathbf{r}_k$ for $k=1,\dots, N_v$ \cite{farhadifar2007}.  Here time has been scaled so that the vertex drag coefficient is unity.  We consider two refinements to this approach.  First, in accounting for substrate drag alone it is natural to assume that the drag force is proportional to a surface area of contact associated with each vertex.  One natural choice for this area, defined solely in terms of the polygonal cell network, is $E_k$, introduced in (\ref{eq:inner0}).  The monolayer is then assumed to evolve under $E_k \dot{\mathbf{r}}_k=-\partial U/\partial \mathbf{r}_k$.

For a vertex in the interior of a monolayer, $E_k$ is composed of three triangles, one lying in each adjacent cell; a vertex at the monolayer periphery, adjacent to one or two triangles, will therefore experience relatively lower drag.  A further feature of this choice is that $(1/E_k) \partial A_i/\partial \mathbf{r}_k$ has a direct interpretation as a discrete approximation of the spatial $\nabla$ operator in $\mathbb{R}^2$ (Appendix~\ref{app:hh}), from which a discrete Laplacian operator (in a Euclidean metric) can be  constructed, acting on scalar fields defined over cells.  

Movement of vertices down energy gradients can then be written (at vertex $k$, or over the whole monolayer) as
\begin{equation}
    E_k\dot{\mathbf{r}_k}=-{\textstyle\sum_\alpha} g_\alpha \mathbf{M}_{\alpha k}, \quad\mathrm{i.e.}\quad \mathsf{E}\dot{\mathsfbfit{r}}=-U_\mathsfbfit{r}\equiv -\mathsfbfit{M}^\top\mathsf{g}, 
    \label{eq:v}
\end{equation}
with $\mathsf{g}$ satisfying (\ref{eq:G}).  We can interpret (\ref{eq:v}) as a force balance on vertices, with drag on the left-hand side balancing elastic forces on the right; in a more global interpretation, (\ref{eq:v}) is a Darcy-type relation expressing the velocity field $\dot{\mathsfbfit{r}}$ as a gradient (in a sense defined below) of pressures and tensions.

A second model of dissipation is relevant when adhesion to the substrate is weak relative to dissipation internal to cells.  The rate of dissipation due to area and perimeter changes can be written (following (\ref{eq:inner0}) and \cite{ANB2018b}) as
\begin{equation}
\Phi\left(\dot{\mathsf{s}}\right)= \left\langle \dot{\mathsf{s}}, \dot{\mathsf{s}} \right\rangle_{\mathsf{H}}\quad\mathrm{where}\quad \mathsf{H}\equiv\left(\begin{matrix}\gamma_A  & {0} \\ {0} & \gamma_L \end{matrix}\right)\otimes \mathsf{I}_{N_c},
\label{eq:inner1}
\end{equation}
for some positive parameters $\gamma_A$ and $\gamma_L$.  Here $\mathsf{I}_{N_c}$ is the $N_c\times N_c$ identity matrix.  Equating the rate of change of mechanical energy with the rate of dissipation, the evolution in this case follows
\begin{equation}
    0=\dot{U}+\Phi=\mathsf{g}^\top \dot{\mathsf{s}}+\dot{\mathsf{s}}^\top \mathsf{H}\dot{\mathsf{s}}=\left(\mathsf{g}^\top \mathsfbfit{M}+\dot{\mathsf{s}}^\top \mathsf{H}\mathsfbfit{M}\right)\cdot\dot{\mathsfbfit{r}},
    \label{eq:dudt}
\end{equation}
indicating that the net force on the vertex (the quantity conjugate to the vertex velocity) is $-\mathsfbfit{M}^\top \mathsf{g}-\mathsfbfit{M}^\top \mathsf{H}\mathsfbfit{M}\cdot \dot{\mathsfbfit{r}}$.  Combining this force with the substrate drag model (\ref{eq:v}) leads to the composite evolution equation
\begin{equation}
\left( \mathsf{E}\otimes \mathsf{I}_2 + \mathsfbfit{M}^\top \mathsf{H}\mathsfbfit{M}\right)\cdot \dot{\mathsfbfit{r}}=-\mathsfbfit{M}^\top\mathsf{g}.
\label{eq:v1}
\end{equation}
Two alternative nonlocal models of dissipation are discussed by \cite{tong2023}, involving relative motion of neighbouring vertices and relative motion of neighbouring cell centres, but we do not pursue these here.  Projecting (\ref{eq:v1}) onto $\dot{\mathsfbfit{r}}$ shows that energy is dissipated as vertices move via
\begin{equation}
    \dot{U}=-\Phi -\dot{\mathsfbfit{r}}^\top \cdot \mathsf{E} \dot{\mathsfbfit{r}} \leq 0.
    \label{eq:diss}
\end{equation}
The evolution (\ref{eq:G}, \ref{eq:v1}) can formally be written as the gradient flow $\dot{\mathsfbfit{r}}=-\mathrm{grad}_{\mathcal{D}}\, U$, using the dissipation operator in (\ref{eq:v1}) as a metric; see Appendix~\ref{app:grfl}.  Here, the gradient is in a function space defined by the dissipation, and it is implemented formally as a $2N_v\times 2N_v$ matrix that can be expected to be dense when $\mathsf{H}$ is non-zero.  

To relate vertex dynamics to cell dynamics, we project (\ref{eq:v1}) onto $\mathsfbfit{M}$ using (\ref{eq:sdot}), to give (\ref{eq:G}) plus
\begin{equation}
\left(\mathsf{I}_{2N_c}+\mathcal{L} \mathsf{H}\right)\dot{\mathsf{s}}=-\mathcal{L}\mathsf{g} \quad \mathrm{where}\quad \mathcal{L}\equiv \mathsfbfit{M}\cdot \mathsf{E}^{-1}\mathsfbfit{M}^\top\equiv \left( \begin{matrix}\mathcal{L}_A & \mathcal{L}_C \\ \mathcal{L}_C^\top & \mathcal{L}_L\end{matrix}\right).
\label{eq:sprob}
\end{equation}
The operator $\mathcal{L}$ combines matrix multiplication (contracting over vertices) and a dot product (contracting over vectors in $\mathbb{R}^2$) to construct a $2\times 2$ matrix of $N_c\times N_c$ blocks with scalar components; diagonal elements $\mathcal{L}_A$ and $\mathcal{L}_L$ describe respectively squared area and perimeter changes of individual cells as a result of vertex displacements; off-diagonal elements $\mathcal{L}_C$ and $\mathcal{L}_C^\top$ describe the magnitude of area and perimeter changes arising from displacements of vertices shared by neighbouring cells. Equivalently, (\ref{eq:sprob}) can be expressed in terms of evolving pressures and tensions using 
\begin{equation}
\left(\mathsf{I}_{2N_c}+\mathsf{G}\mathcal{L} \mathsf{H}\mathsf{G}^{-1}\right)\dot{\mathsf{g}}=-\mathsf{G}\mathcal{L} \mathsf{g}.
\label{eq:g}
\end{equation}

Typically, pressures and tensions are coupled via the off-diagonal blocks $\mathcal{L}_C$ and $\mathcal{L}_C^\top$ of $\mathcal{L}$.  For extreme values of $\Gamma$, however (see (\ref{eq:nonlinG})), the operator $\mathsf{G}\mathcal{L}$ decouples into two components (Appendix~\ref{app:gam}) representing fast cell dilation and slow cell shear.  For small (large) $\Gamma$, cell dilation is driven predominantly by pressure (tension) changes, and the relevant component of (\ref{eq:g}) is approximated by
\begin{subequations}
\begin{align}
    (\mathsf{I}+\gamma_A \mathsf{A}_c^{-1}\mathcal{L}_A \mathsf{A}_c )\dot{\mathsf{P}}&=-\mathsf{A}_c^{-1}\mathcal{L}_A \mathsf{P}, \quad \quad \quad (\Gamma\sim \gamma_L\ll 1) \\ 
    (\mathsf{I}+\gamma_L \mathsf{L}_{c}^{-1}\mathcal{L}_L \mathsf{L}_{c})\dot{\mathsf{T}}&=-\Gamma L_0\mathsf{L}_{c}^{-1}\mathcal{L}_L \mathsf{T}, \quad~(\Gamma\sim \gamma_A^{-1}\gg 1),
\end{align}
    \label{eq:dilation}
\end{subequations}
where $\mathsf{P}\equiv (P_1, \dots, P_{N_c})$ and $\mathsf{T}\equiv (T_1,\dots,T_{N_c})$; $\sim$ denotes `scales like'.  The operator $-\mathsf{A}_c^{-1}\mathcal{L}_A$ in (\ref{eq:dilation}a), the first diagonal block of $\mathsf{G}\mathcal{L}$, is a discrete approximation of the spatial $\nabla^2$ operator in $\mathbb{R}^2$ (Appendix~\ref{app:hh}) when approximated on the polygonal grid defined by cells using the metrics ${\mathsf{E}}$ and ${\mathsf{G}}$ (see (\ref{eq:nonlinG}, \ref{eq:inner0})), indicating that cell pressures can exhibit diffusive dynamics (mediated by viscous resistance to area change that is proportional to $\gamma_A$).  The corresponding operator $-\mathsf{L}_{c}^{-1}\mathcal{L}_L$ in (\ref{eq:dilation}b) does not appear to have such a straightforward interpretation in terms of conventional derivatives.

In summary, (\ref{eq:G}, \ref{eq:v1}) provides a full description of the dynamics (with $2N_v$ degrees of freedom), while (\ref{eq:sprob}) and (\ref{eq:g}) each describe the evolution of $2N_c$ cell-based variables, albeit depending on the $2N_v\times 2N_v$ matrix $\mathcal{L}$.  Despite a suggestion of diffusive dynamics in (\ref{eq:g}, \ref{eq:dilation}), this interpretation is imperfect because the operator $\mathcal{L}$ evolves with $\mathsf{s}$ and $\mathsf{g}$ via its dependence on $\mathsfbfit{r}$; this can have a significant influence on the dynamics of the system.

\subsection{Cell stress}

At equilibrium, $\mathsf{g}$ and ${\mathsf{s}}$ are typically non-uniform (with $\mathsf{g}$ in the kernel of $\mathsfbfit{M}^\top$), representing the fact that there can be residual stress (or prestress) in the equilibrium state of a disordered monolayer.  The stress of cell $i$ is $\boldsymbol{\sigma}_i\equiv \mathsf{I}_2 P_i+(L_iT_i/A_i)\mathsfbfit{Q}_i$, where $\mathsfbfit{Q}_i$ is a shape tensor \cite{ANB2018b, ANB2018a} defined in (\ref{eq:firstmoment}b) (Appendix~\ref{sec:appx}) satisfying $\mathrm{Tr}(\mathsfbfit{Q}_i)=1$; it quantifies the degree to which a cell is sheared. 
We can characterise cell stress through the magnitudes of the isotropic and deviatoric components of $\boldsymbol{\sigma}_i$ \cite{ANB2018a,jensen2020}, namely
\begin{equation}
    P_{\mathrm{eff},i}=P_i+\frac{L_iT_i}{2A_i}, \quad \zeta_i= \frac{L_i T_i}{A_1} \sqrt{-\mathrm{det}(\mathsf{Q}_i-\tfrac{1}{2}\mathsf{I}_2)}.
    \label{eq:peffshear}
\end{equation}
An isolated hexagonal cell, for which $P_{\mathrm{eff},i}=0$ and $\mathsf{Q}_i=\tfrac{1}{2}\mathsf{I}_2$, can still exhibit prestress at the level of individual vertices when $T_i>0$ (\hbox{i.e.} for $L_0\lesssim 3.72$).

Defining $\mathsfbfit{W}\equiv \mathrm{diag}(\mathsf{I}_{2},\dots,\mathsf{I}_2,\mathsfbfit{Q}_1,\dots,\mathsfbfit{Q}_{N_c})$, (\ref{eq:firstmoment}) shows that
$\mathsfbfit{r}\otimes \mathsfbfit{M}^\top =\mathsfbfit{W}\mathsf{s}$ 
and hence, taking the trace,
\begin{equation} 
\mathsfbfit{M}\cdot\mathsfbfit{r}=(2A_1,\dots,2A_{N_c}, L_1,\dots ,L_{N_c})^\top;
\label{eq:mdiv}
\end{equation}
(the factor of 2 is analogous to the identity $\nabla\cdot\mathbf{x}=2$ in $\mathbb{R}^2$), again hinting at the divergence-like nature of $\mathsfbfit{M}\cdot$.  Applying $\mathsfbfit{r}\,\otimes$ to the force balance equation (\ref{eq:v1}) leads to
\begin{equation}
    \frac{\mathrm{d}}{\mathrm{d}t}\left\{\tfrac{1}{2} \mathsfbfit{r}\otimes \mathsf{E}\mathsfbfit{r} + \tfrac{1}{2} \mathsf{s}^\top \mathsfbfit{W} \mathsf{H}\mathsf{s}\right\}=-\mathsf{s}^\top \mathsfbfit{W} \mathsf{g}\equiv -{\textstyle{\sum_i}} A_i \boldsymbol{\sigma}_i.
    \label{eq:stressevol}
\end{equation}
This sum over cells equates the elastic cell stress $\boldsymbol{\sigma}_i$, integrated over the monolayer, with expressions measuring the rate of working of dissipative forces associated with vertex motion and changes in area and perimeter.  At equilibrium, both sides of (\ref{eq:stressevol}) tend to zero, although the stress in individual cells can be non-uniform.

As shown in Eq.~(23) of \cite{ANB2018b}, the stiffness of a monolayer (under imposition of a small affine deformation with strain $\mathfrak{E}$) includes terms involving $\mathrm{Tr}(\mathfrak{E})$ and  $\mathsfbfit{Q}_i:\mathfrak{E}$ 
(associated with changes in cell areas and perimeters respectively) and $\mathsfbfit{B}_i:\mathfrak{E}$ (where $\mathsfbfit{B}_i$ is a fourth-order tensor, describing changes in $\mathsfbfit{Q}_i$ via reorientation of edges).  The former terms contribute to the so-called material stiffness of the monolayer (by changing the energy directly), while the latter contributes to its geometric stiffness (providing stiffness without directly changing areas and perimeters).  We now consider the role of these different forms of stiffness in the viscoelastic relaxation of a monolayer.

\subsection{Linearized dynamics}

To better understand the relationship between the evolution of vertices and cells, we consider how the system relaxes to an equilibrium state.  We write $\mathsf{s}=\bar{\mathsf{s}}+\hat{\mathsf{s}}(t)$ (with $\langle\hat{\mathsf{s}},\hat{\mathsf{s}}\rangle_{\mathsf{G}}\ll\langle\bar{\mathsf{s}},\bar{\mathsf{s}}\rangle_{\mathsf{G}}$, etc.) to distinguish a steady state (with a bar) from decaying perturbations (with hats).  In force balances, we neglect terms that are quadratic (or smaller) in hatted quantities.  
We Taylor-expand $\mathsfbfit{M}$ with respect to $\mathbf{r}_k$ using (\ref{eq:x}) to obtain $\hat{\mathsfbfit{M}}=\bar{\mathsfbfit{M}'}\cdot \hat{\mathsfbfit{r}}$, so that $\hat{\mathsfbfit{M}}^\top=\hat{\mathsfbfit{r}}^\top \cdot \bar{\mathsfbfit{M}'}^\top$.  Steady states satisfy
\begin{equation} 
\bar{\mathsf{g}}=\mathcal{G}(\bar{\mathsf{s}};\mathsf{s}_0,\Gamma),     \quad
    \bar{\mathsfbfit{M}}^\top\bar{\mathsf{g}}=\mathsf{0}. 
    \label{eq:lineq1}
\end{equation}

We then assume that all hatted quantities have time dependence $e^{-\lambda t}$, seeking the spectrum $\lambda^{(1)},\dots,\lambda^{(2N_V)}$ defined by possible values of $\lambda$; we suppress the superscript in what follows.  Given the gradient-flow structure of the problem, we expect $\lambda\in \mathbb{R}$ with $\lambda\geq 0$.  Evolving states satisfy, from (\ref{eq:nonlinG}, \ref{eq:sdot}, \ref{eq:v1}),
\begin{align}    
\hat{\mathsf{g}}=\bar{\mathsf{G}}\hat{\mathsf{s}}, \quad \hat{\mathsf{s}}=\bar{\mathsfbfit{M}}\cdot \hat{\mathbf{r}}, \quad -\lambda \left(\bar{\mathsf{E}}\otimes \mathsf{I}_2 + \bar{\mathsfbfit{M}}^\top \mathsf{H}\bar{\mathsfbfit{M}}\right)\cdot\hat{\mathbf{r}}=-\bar{\mathsfbfit{M}}^\top\hat{\mathsf{g}}-\hat{\mathsfbfit{M}}^\top\bar{\mathsf{g}}.
\end{align}
 
Thus, using $\hat{\mathcal{L}}=\hat{\mathsfbfit{M}}\cdot \bar{\mathsf{E}}^{-1} \bar{\mathsfbfit{M}}^\top+\bar{\mathsfbfit{M}}\cdot \hat{\mathsf{E}}^{-1} \bar{\mathsfbfit{M}}^\top+\bar{\mathsfbfit{M}}\cdot \bar{\mathsf{E}}^{-1} \hat{\mathsfbfit{M}}^\top$, the linearized dynamics of vertices and cells follows, respectively, 
\begin{subequations}
\begin{align}  
 \lambda \left( \bar{\mathsf{E}}\otimes \mathsf{I}_2 + \bar{\mathsfbfit{M}}^\top \mathsf{H}\bar{\mathsfbfit{M}}\right)\cdot\hat{\mathsfbfit{r}} &=   \left( \bar{\mathsfbfit{M}}^\top \bar{\mathsf{G}} \bar{\mathsfbfit{M}} +\bar{\mathsf{g}}^\top \bar{\mathsfbfit{M}'} \right) \cdot \hat{\mathsfbfit{r}},\\
\lambda \left(\mathsf{I}_{2N_c}+\bar{\mathcal{L}}\mathsf{H}\right)\hat{\mathsf{s}}& =\bar{\mathcal{L}}\bar{\mathsf{G}}\hat{\mathsf{s}}+\hat{\mathcal{L}}\bar{\mathsf{g}}
=\bar{\mathcal{L}}\bar{\mathsf{G}}\hat{\mathsf{s}}+\bar{\mathsfbfit{M}}\cdot \bar{\mathsf{E}}^{-1} (\hat{\mathsfbfit{r}}^\top \cdot\bar{\mathsfbfit{M}'}^\top)\bar{\mathsf{g}}.
\end{align}
\label{eq:lin}
\end{subequations}
We recognise $ \bar{\mathsfbfit{M}}^\top \bar{\mathsf{G}} \bar{\mathsfbfit{M}} +\bar{\mathsf{g}}^\top \bar{\mathsfbfit{M}'}$ in (\ref{eq:lin}a) as the Hessian $\bar{U}_{\mathsfbfit{r}\mathsfbfit{r}}$, with terms representing respectively material stiffness and geometric stiffness \cite{guest2006, damavandi2022}.  Equivalently, expanding the energy to second order with respect to vertex locations using (\ref{eq:x}) gives
\begin{equation}
    U\approx {\textstyle \sum_i} \bar{U}_i
    +\hat{\mathsfbfit{r}}^\top\cdot \bar{\mathsfbfit{M}}^\top \bar{\mathsf{g}}+\tfrac{1}{2}\hat{\mathsfbfit{r}}^\top\cdot\left[\bar{\mathsfbfit{M}}^\top\bar{\mathsf{G}}\bar{\mathsfbfit{M}}+\bar{\mathsf{g}}^\top\bar{\mathsfbfit{M}'}\right]\cdot\hat{\mathsfbfit{r}}.
    \label{eq:energyvar}
\end{equation}
The evolution of cell pressures and areas is coupled to movement of vertices in (\ref{eq:lin}b) via $\bar{\mathsfbfit{M}'}^\top \bar{\mathsf{g}}$.  A geometric interpretation of the associated forces is given in Appendix~\ref{sec:prst} and steps needed to evaluate $\bar{\mathsfbfit{M}}$ and $\bar{\mathsfbfit{M}}'$ are outlined in Appendix~\ref{sec:appx}.  The scalar and vector operators $\bar{\mathcal{L}}$ and $\bar{\mathsfbfit{M}}^\top\bar{\mathsf{G}}\bar{\mathsfbfit{M}}\cdot$ are suggestive of Laplacian operators, in a sense that we will qualify below.  
The relaxation rates $\lambda$ appear as eigenvalues of the generalized eigenvalue problem (\ref{eq:lin}a). 
The $\lambda$-spectrum was evaluated by \cite{tong2023} and \cite{damavandi2022B} under periodic boundary conditions (using different models of dissipation).  Here, we seek an understanding of the contributions of the Laplacian operators in (\ref{eq:lin}) to the dynamics of an isolated monolayer.

\subsection{Simulations}

We simulated the growth of isolated 100-cell monolayers, implementing a stochastic cell-division algorithm and T1 transitions to produce realisations of disordered monolayers for chosen values of $\Gamma$ and $L_0$; the monolayer periphery is force-free. Monolayers are successively relaxed to equilibrium for different parameter choices. We chose a disordered monolayer such that it maintained the same topology across the parameter space and did not contain any quadrilaterals at equilibrium. (Quadrilaterals can add the additional complexity of retaining localised prestress beyond the usual rigidity transition because their shape incompatability does not allow them to relax their tension; in the jammed regime the exclusion of quadrilaterals makes a negligible difference to results.) We also simulated a symmetric 127-cell monolayer of hexagonal cells. In what follows, the parameters $\Gamma$ and $L_0$ are chosen primarily so that the monolayer is in a jammed state. 

Simulation code is available as the \texttt{VertexModel.jl} package~\cite{Revell_VertexModel_jl_2022}, previously developed for our past work in this area~\cite{jensen2022}. The \texttt{DiscreteCalculus.jl} package~\cite{Revell_DiscreteCalculus_jl} was developed to aid the analysis in this paper. This package contains Julia code that creates operators discussed in this paper. 

\section{Singular-value decomposition}
\label{sec:svd}

\begin{figure}
\begin{center}
    \includegraphics[height=1.7in]{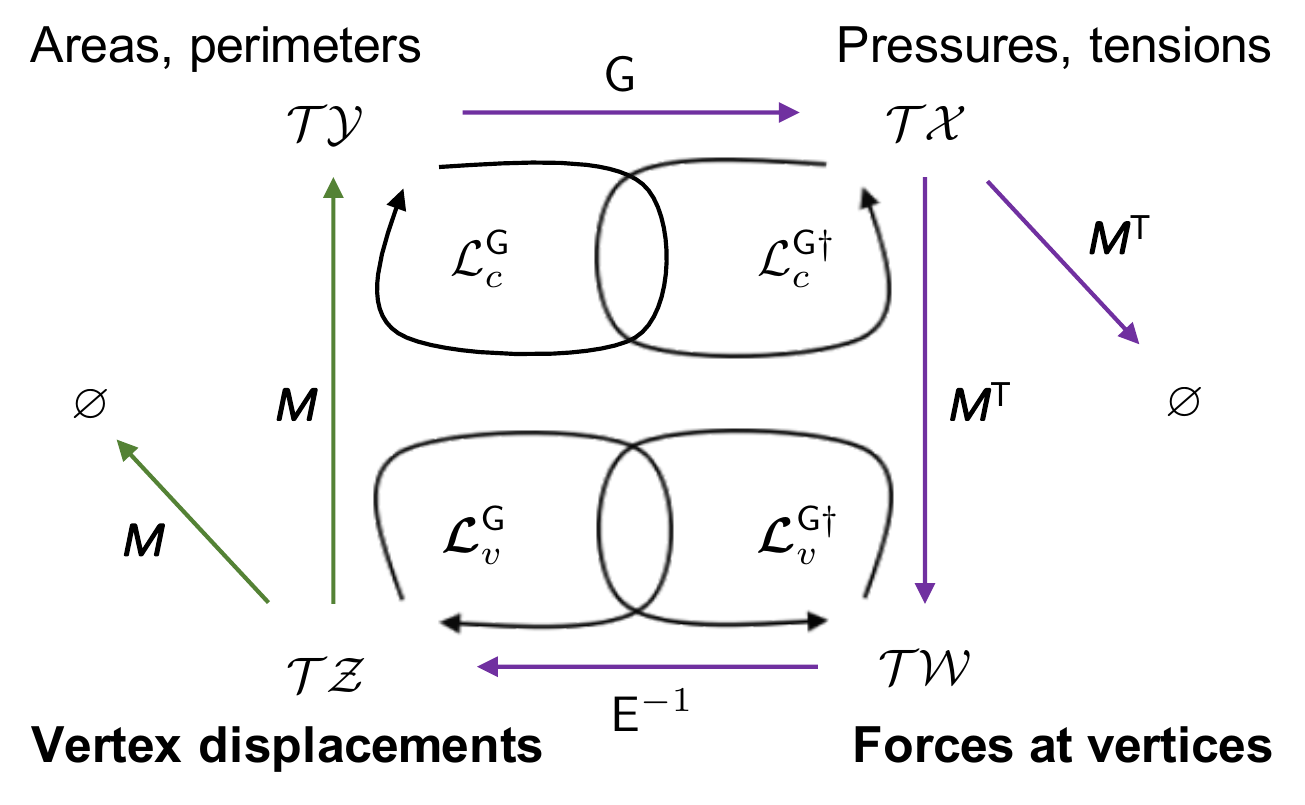}~~~~~~~~~
    \includegraphics[height=1.7in]{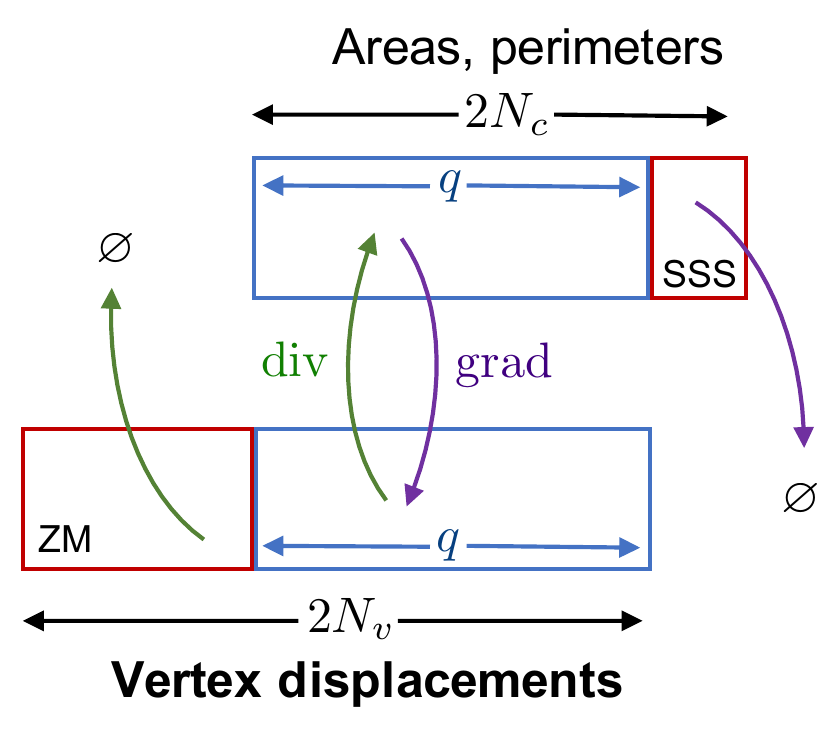}
\end{center}
    \caption{Left: a diagram summarising the action of matrix operators $\mathsfbfit{M}$, $\mathsf{G}$ and $\mathsf{E}$.   ${\mathsfbfit{M}}$ has components that are vectors describing how cell areas and perimeters change under small vertex displacements. `Vertex displacements' describes the tangent space $\mathcal{T}\mathcal{Z}$ of the state space $\mathcal{Z}$.  `Forces at vertices' describes the dual space $\mathcal{T}\mathcal{W}$; `Pressures, tensions' sit in a space $\mathcal{TX}$ that is dual to the tangent space $\mathcal{TY}$ denoted `Areas, perimeters.'  The operators $\mathsfbfit{M}\cdot$ and $\mathsfbfit{M}^\top$ are singular with rank $q<2N_c$ and therefore have nontrivial kernels mapping to $\varnothing$.  Laplacians $\mathcal{L}_c^{\mathsf{G}}$ and $\boldsymbol{\mathcal{L}}_v^{\mathsf{G}}$ are defined in (\ref{eq:lapcv}); dual operators are given in (\ref{eq:lapcvdual}).  Right: kernels in $\mathcal{TZ}$ and $\mathcal{TY}$ are shown as red boxes.  Perturbations are mapped between these spaces by divergence $\langle\mathsfbfit{N}^\top, \cdot \rangle_{\mathsf{E}}$ and gradient $\langle\mathsfbfit{N}, \cdot \rangle_{\mathsf{G}}$ matrix operators.  Elements of $\mathrm{ker}(\bar{\mathsfbfit{M}}^\top)$ are denoted States of Self-Stress (SSS); elements of $\mathrm{ker}(\bar{\mathsfbfit{M}}\cdot)$ are denoted Zero Modes (ZM).}
    \label{fig:svdsummary}
\end{figure}

In order to relate the full spectrum describing relaxation of vertices (with $2N_v$ eigenvalues) to the potentially smaller system describing the evolution of $2N_c$ scalars defined over cells, it is helpful to reframe the problem using singular-value decomposition (SVD).  SVD provides a framework for understanding the action of the singular matrix operators $\mathsfbfit{M}$ and $\mathsfbfit{M}^\top$, by identifying associated Laplacian operators (illustrated in Figure~\ref{fig:svdsummary}) and enabling an assessment of their contribution to the overall dynamics.

Let $\mathcal{Z}\subset\mathbb{R}^{N_v}\times \mathbb{R}^2$ be the (state) space of vertex locations and $\mathcal{T}\mathcal{Z}$ be its tangent space.  Let $\mathcal{Y}\subset\mathbb{R}^{2N_c}$ be the space of areas and perimeters and $\mathcal{T}\mathcal{Y}$ be its tangent space.  The matrix operator $\mathsfbfit{M}\cdot$ (see (\ref{eq:x})) maps from $\mathcal{T}\mathcal{Z}$ to $\mathcal{T}\mathcal{Y}$ (Fig.~\ref{fig:svdsummary}).  We define $q$ such that $\mathrm{rank}(\mathsfbfit{M})\equiv q < 2N_c<2N_v$.  For a scalar-valued field $\mathsf{Y}\in \mathcal{T}\mathcal{Y}$, $\mathsf{G}\mathsf{Y}$ represents the associated variations in pressures and tensions; the norm $\tfrac{1}{2}\langle \mathsf{Y},\mathsf{Y} \rangle_{\mathsf{G}}$ (see (\ref{eq:secondenergy})) measures the change in elastic energy associated with area and perimeter variation.  $\mathsf{GY}$ sits in $\mathcal{TX}$, the tangent space to the space of pressures and tensions, which is dual to $\mathcal{TY}$ under (\ref{eq:inner0}).  For $\mathsfbfit{Z}\in \mathcal{TZ}$, the norm $\tfrac{1}{2}\langle \mathsfbfit{Z},\mathsfbfit{Z} \rangle_{\mathsf{E}}$ measures the dissipation associated with vertex motion.  $\mathcal{TW}$ (the space of force variations) is dual to $\mathcal{TZ}$ under (\ref{eq:inner0}).

We then define $\mathsfbfit{N}=\mathsfbfit{M}\mathsf{E}^{-1}$ 
so that $\langle\mathsfbfit{N}^\top,\cdot \rangle_{\mathsf{E}}$ maps from $\mathcal{TZ}$ to $\mathcal{TY}$.  Exploiting SVD \cite{pellegrino1993}, we can express $\mathsfbfit{N}$ acting under the inner products (\ref{eq:inner0}) as
\begin{equation}
    \mathsfbfit{N}=
    \sum_{p=1}^{q} \sigma_p \mathsf{Y}_p \mathsfbfit{Z}_p^\top\quad \mathrm{where}\quad \begin{cases}
        \langle\mathsfbfit{Z}_p,\mathsfbfit{Z}_{p'}\rangle_{\mathsf{E}}=\delta_{p p'}& (1\leq p,p' \leq N_v), \\
         \langle\mathsf{Y}_p,\mathsf{Y}_{p'}\rangle_{\mathsf{G}}=\delta_{pp'}& (1\leq p,p' \leq 2N_c).
    \end{cases} 
    \label{eq:svd}
\end{equation}
Here $\left\{\mathsf{Y}_1\dots\mathsf{Y}_{2N_c}\right\}$ form an orthonormal basis of $\mathcal{TY}$ and are the left-singular vectors of $\mathsfbfit{N}$; $\left\{\mathsfbfit{Z}_1,\dots , \mathsfbfit{Z}_{N_v}  \right\}$ form an orthonormal basis of $\mathcal{TZ}$ and are the right-singular vectors of $\mathsfbfit{N}$.  The non-zero singular values of $\mathsfbfit{N}$ are ordered as $0< \sigma_1\leq \sigma_2\leq \dots \leq \sigma_q$ and the singular vectors are related via 
\begin{subequations}
    \begin{align}
    \langle \mathsfbfit{N}^\top,\mathsfbfit{Z}_p\rangle_{\mathsf{E}} &=\sigma_p\mathsf{Y}_p \quad (1\leq p\leq q),& \langle\mathsfbfit{N}, \mathsf{Y}_p\rangle_{\mathsf{G}}&=\sigma_p \mathsfbfit{Z}_p \quad (1\leq p\leq q), \label{eq:nonzeromodes} \\ \langle \mathsfbfit{N}^\top, \mathsfbfit{Z}_p\rangle_{\mathsf{E}} &=\mathsf{0} \quad (q+1\leq p\leq 2N_v), & \langle\mathsfbfit{N} ,\mathsf{Y}_p\rangle_{\mathsf{G}}&=\mathsf{0}\quad (q+1\leq p\leq 2N_c). 
\end{align}
\label{eq:zeromodes}
\end{subequations}
From this it follows that 
\begin{align}
\langle \mathsfbfit{N}^\top,\mathsfbfit{N}^\top\rangle_{\mathsf{E}}&=    \mathsfbfit{N}\mathsf{E}\cdot \mathsfbfit{N}^\top =\sum_{p=1}^q \sigma_p^2 \mathsf{Y}_p \mathsf{Y}_p^\top,&
\langle\mathsfbfit{N},\mathsfbfit{N}\rangle_{\mathsf{G}}=
    \mathsfbfit{N}^\top \mathsf{G}\mathsfbfit{N} &=\sum_{p=1}^q \sigma_p^2 \mathsfbfit{Z}_p \mathsfbfit{Z}_p^\top.
\end{align}
Thus, defining the vertex and cell operators
\begin{align}
    \boldsymbol{\mathcal{L}}^{\mathsf{G}}_v&=\mathsfbfit{N}^\top \mathsf{G} \mathsfbfit{N} \mathsf{E}\cdot\,=\mathsf{E}^{-1}\mathsfbfit{M}^\top \mathsf{G} \mathsfbfit{M} \cdot\,, &
    \mathcal{L}^{\mathsf{G}}_c&=\mathsfbfit{N} \mathsf{E} \cdot\mathsfbfit{N}^\top \mathsf{G}=\mathsfbfit{M} \mathsf{E}^{-1} \cdot\mathsfbfit{M}^\top \mathsf{G}\equiv \mathcal{L}\mathsf{G},
    \label{eq:lapcv}
\end{align}
it follows that, for $1\leq p\leq q$, singular values can be related to eigenvalues of (\ref{eq:lapcv}) via
\begin{align}
    \langle\mathsfbfit{N},\langle \mathsfbfit{N}^\top,\mathsfbfit{Z}_p\rangle_{\mathsf{E}} \rangle_{\mathsf{G}}&\equiv 
    \boldsymbol{\mathcal{L}}^{\mathsf{G}}_v
    \mathsfbfit{Z}_p= \sigma_p^2 \mathsfbfit{Z}_p, &
    \langle\mathsfbfit{N}^\top ,\langle\mathsfbfit{N},\mathsf{Y}_p\rangle_{\mathsf{G}}\rangle_{\mathsf{E}} &\equiv \mathcal{L}^{\mathsf{G}}_c
    \mathsf{Y}_p=\sigma_p^2 \mathsf{Y}_p. 
    \label{eq:19}
\end{align}
The operators $\boldsymbol{\mathcal{L}}_v^{\mathsf{G}}$ and $\mathcal{L}_c^{\mathsf{G}}$ are self-adjoint under the relevant inner products, because (taking transposes, for any fields $\mathsfbfit{Z}$, $\tilde{\mathsfbfit{Z}}$, $\tilde{\mathsf{Y}}$, $\mathsf{Y}$)
\begin{subequations}
\begin{gather}
    \langle \tilde{\mathsfbfit{Z}},\boldsymbol{\mathcal{L}}^{\mathsf{G}}_v \mathsfbfit{Z} \rangle_{\mathsf{E}}=\tilde{\mathsfbfit{Z}}^\top\cdot \mathsfbfit{M}^\top \mathsf{G}\mathsfbfit{M} \cdot \mathsfbfit{Z}=
    \mathsfbfit{Z}^\top\cdot \mathsfbfit{M}^\top \mathsf{G}\mathsfbfit{M} \cdot \tilde{\mathsfbfit{Z}}= \langle \mathsfbfit{Z},\boldsymbol{\mathcal{L}}^{\mathsf{G}}_v \tilde{\mathsfbfit{Z}} \rangle_{\mathsf{E}},\\
        \langle \tilde{\mathsf{Y}},\mathcal{L}^{\mathsf{G}}_c \mathsf{Y} \rangle_{\mathsf{G}}=\tilde{\mathsf{Y}}^\top \mathsf{G}\mathsfbfit{M} \mathsf{E}^{-1}\cdot \mathsfbfit{M}^\top \mathsf{G} \mathsf{Y}=
    \mathsf{Y}^\top \mathsf{G}\mathsfbfit{M} \mathsf{E}^{-1} \cdot \mathsfbfit{M}^\top \mathsf{G}  \tilde{\mathsf{Y}}= \langle \mathsf{Y},\mathcal{L}^{\mathsf{G}}_c \tilde{\mathsf{Y}} \rangle_{\mathsf{G}}.
\end{gather}
\end{subequations}
In addition, the operators are positive semi-definite because, for any $\mathsfbfit{Z}$, $\mathsf{Y}$, 
\begin{align}
    \langle {\mathsfbfit{Z}}, \boldsymbol{\mathcal{L}}^{\mathsf{G}}_v \mathsfbfit{Z} \rangle_{\mathsf{E}}&=\langle \langle \mathsfbfit{N}^\top, \mathsfbfit{Z}\rangle_{\mathsf{E}}, \langle\mathsfbfit{N}^\top , \mathsfbfit{Z}\rangle_{\mathsf{E}} \rangle_{\mathsf{G}} \geq 0,  &
        \langle{\mathsf{Y}},\mathcal{L}^{\mathsf{G}}_c \mathsf{Y} \rangle_{\mathsf{E}}& =\langle  \langle \mathsfbfit{N}, \mathsf{Y}\rangle_{\mathsf{G}},  \langle \mathsfbfit{N} , \mathsf{Y}\rangle_{\mathsf{G}} \rangle_{\mathsf{E}} \geq 0.
\end{align}
The operators also satisfy
 $   \mathsfbfit{N}\mathsf{E}\cdot \boldsymbol{\mathcal{L}}^{\mathsf{G}}_v=\mathcal{L}^{\mathsf{G}}_c \mathsfbfit{N}\mathsf{E}\cdot$ and  
$    \mathsfbfit{N}^\top\mathsf{G} \mathcal{L}^{\mathsf{G}}_c=\boldsymbol{\mathcal{L}}^{\mathsf{G}}_v \mathsfbfit{N}^\top\mathsf{G}$.
In summary, as illustrated in  Fig.~\ref{fig:svdsummary}, the gradient operator $\langle \mathsfbfit{N},\cdot\rangle_{\mathsf{G}}$ (the matrix operator $\mathsf{E}^{-1}\mathsfbfit{M}^\top \mathsf{G}$) maps areas and perimeters to vertex displacements; the divergence operator $\langle \mathsfbfit{N}^\top, \cdot \rangle_{\mathsf{E}}$ (the matrix operator $\mathsfbfit{M}^\top\cdot$) sums vertex displacements over cells to give changes in areas and perimeters.  The combinations div~$\circ$~grad and grad~$\circ$~div in (\ref{eq:lapcv}) form scalar (cell) and vector (vertex) Laplacians $\mathcal{L}_c^{\mathsf{G}}$ and $\boldsymbol{\mathcal{L}}_v^{\mathsf{G}}$ respectively.  The left- and right-singular vectors are the eigenvectors of $\mathcal{L}_c^{\mathsf{G}}$ and $\boldsymbol{\mathcal{L}}_v^{\mathsf{G}}$ respectively; the two families of eigenmodes can be related to each other via (\ref{eq:nonzeromodes}).  The zero modes of the two operators (\ref{eq:zeromodes}) are not directly related to each other (for example, translation and rotation modes will be among the zero modes of $\boldsymbol{\mathcal{L}}_v^{\mathsf{G}}$ but do not have an analogue in $\mathcal{L}_c^{\mathsf{G}}$).  For a monolayer with symmetries, we can expect some singular values to be repeated.

Equivalently, we can also define Laplacians associated with dissipation due to area and perimeter changes
\begin{align}
    \boldsymbol{\mathcal{L}}_v^{\mathsf{H}}&=\mathsfbfit{N}^\top \mathsf{H}\mathsfbfit{N}\mathsf{E}\cdot=\mathsf{E}^{-1}\mathsfbfit{M}^\top \mathsf{H} \mathsfbfit{M} \cdot\,, &
    \mathcal{L}_c^{\mathsf{H}}&=\mathsfbfit{N}\mathsf{E}\cdot\mathsfbfit{N}^\top \mathsf{H}=\mathsfbfit{M} \mathsf{E}^{-1} \cdot\mathsfbfit{M}^\top \mathsf{H},
    \label{eq:lapcvh}
\end{align}
which again will share non-zero eigenvalues, and which satisfy $\mathsfbfit{N}\mathsf{E}\cdot \boldsymbol{\mathcal{L}}^{\mathsf{H}}_v=\mathcal{L}^{\mathsf{H}}_c \mathsfbfit{N}\mathsf{E}\cdot$ and $\mathsfbfit{N}^\top\mathsf{H} \mathcal{L}^{\mathsf{H}}_c=\boldsymbol{\mathcal{L}}^{\mathsf{H}}_v \mathsfbfit{N}^\top\mathsf{H}$.

A dual formulation using $\mathsfbfit{N}^\dag\equiv \mathsfbfit{M}^\top\mathsf{G}=\sum_{p=1}^q\sigma_p \mathsfbfit{W}_p \mathsf{X}_p^\top$ maps pressure and tension variations $\mathsf{X}_p\in \mathcal{TX}$ to force variations $\mathsfbfit{W}_p\in \mathcal{TW}$, via the gradient $\langle\mathsfbfit{N}^{\dag\top}, \cdot\rangle_{\mathsf{G}^{-1}}={\mathsfbfit{M}}^\top$ and divergence $\langle\mathsfbfit{N}^\dag,\cdot \rangle_{\mathsf{E^{-1}}}=\mathsf{G}{\mathsfbfit{M}}\mathsf{E}^{-1}\cdot$, yielding Laplacians acting respectively on forces and pressures plus tensions (see Fig.~\ref{fig:svdsummary}):
\begin{align}
    \boldsymbol{\mathcal{L}}^{\mathsf{G}\dag}_v&=\mathsfbfit{N}^\dag \mathsf{G}^{-1} \mathsfbfit{N}^{\dag\top} \mathsf{E}^{-1}\cdot\,=\mathsfbfit{M}^\top \mathsf{G} \mathsfbfit{M}\mathsf{E}^{-1} \cdot\,, &
    \mathcal{L}^{\mathsf{G}\dag}_c&=\mathsfbfit{N}^{\dag\top} \mathsf{E}^{-1} \cdot\mathsfbfit{N}^\dag \mathsf{G}^{-1}=\mathsf{G}\mathsfbfit{M} \mathsf{E}^{-1} \cdot\mathsfbfit{M}^\top \equiv \mathsf{G}\mathcal{L}.
    \label{eq:lapcvdual}
\end{align}
The evolution equation for the monolayer (\ref{eq:v1}) can then be written in terms of forces $\mathsfbfit{f}$ (satisfying $\dot{\mathsfbfit{f}}=\mathsfbfit{E}\dot{\mathsfbfit{r}}$) as
\begin{subequations}
\label{eq:dvosvo}
\begin{equation}
    \dot{\mathsfbfit{f}}+\boldsymbol{\mathcal{L}}^{\mathsf{H}\dag}_v \dot{\mathsfbfit{f}}=-
   \langle \mathsfbfit{N}^{\dag \top}, \mathsf{g}\rangle_{\mathsf{G}^{-1}},
\label{eq:dvo}
\end{equation}
again illustrating the gradient-flow structure of the evolution. 

Applying $\langle\mathsfbfit{N}^\dag, \cdot \rangle_{\mathsf{E}^{-1}}$ to (\ref{eq:dvo}) recovers (\ref{eq:sprob}) in its dual form
\begin{equation}
    \dot{\mathsf{g}}+\mathsf{G}\mathcal{L}^{\mathsf{H}}_c \mathsf{G}^{-1} \dot{\mathsf{g}}=-\mathcal{L}_c^{\mathsf{G}\dag} \mathsf{g}.
    \label{eq:svo}
\end{equation}
\end{subequations}
Pressures and tensions evolve under the operator $\mathcal{L}^{\mathsf{G}\dag}_c$, which has $q$ degrees of freedom but which is determined by the larger number ($2N_v$) of vertices.  Eq.~(\ref{eq:dvo}) generalises to 2D the 1D Lagrangian evolution equation derived in \cite{fozard2010}: the evolving force field (dual to the velocity field) balances a pressure (and tension) gradient plus a viscous term penalising stretching; the associated kinematic relation (\ref{eq:svo}) suggests diffusive dynamics.  However we must again recognise that area and pressure changes can arise from dynamic changes in the operator $\mathcal{L}_c^{\mathsf{G}\dag}$.   

\subsection{Spectral contributions of geometric and material stiffness}

Having set the model in an SVD framework, we now identify the contributions of Laplacians to the overall dynamics.  An equilibrium state (denoted with a bar) satisfies $\bar{\mathsfbfit{M}}^\top \bar{\mathsf{g}}=0$ (see (\ref{eq:lineq1})), placing $\bar{\mathsf{G}}^{-1}\bar{\mathsf{g}}$ in the kernel of $\langle \bar{\mathsfbfit{N}},\cdot \rangle_{\bar{\mathsf{G}}}$ and making $\bar{\mathsf{g}}$ a harmonic function with respect to $\mathcal{L}_c^{\bar{\mathsf{G}\dag}}$. Thus 
\begin{equation}
    \bar{\mathsf{G}}^{-1}\bar{\mathsf{g}}={\textstyle\sum_{p=q+1}^{2N_c}}\gamma_p\bar{\mathsf{Y}}_p
    \label{eq:bars}
\end{equation}
for some coefficients $\gamma_p$.

We require that $q\leq 2N_c-1$, to allow the existence of at least one zero mode of $\bar{\mathcal{L}}^{{\mathsf{G}}}_c$ (a state of self-stress) that can accommodate an equilibrium solution.  

Consider small disturbances 

$\hat{\mathsfbfit{r}}$ and $\hat{\mathsf{s}}=\langle \bar{\mathsfbfit{N}}^\top, \hat{\mathsfbfit{r}} \rangle_{\bar{\mathsf{E}}}$ to the equilibrium.  We Taylor expand $\mathsfbfit{N}$ so that 
$
    \mathsfbfit{N}=\bar{\mathsfbfit{N}}+\langle \bar{\mathsfbfit{N}}',\hat{\mathsfbfit{r}} \rangle_{\bar{\mathsf{E}}}+\dots
$
where $\bar{\mathsfbfit{N}}'=\bar{\mathsf{E}}^{-1}\mathsfbfit{M}'\bar{\mathsf{E}}^{-1}$.  

Assuming dissipation via the substrate dominates that in the cells, (\ref{eq:lineq1}) and (\ref{eq:lin}) become 
\begin{subequations}
\begin{align}
0&=\langle \bar{\mathsfbfit{N}},\bar{\mathsf{G}}^{-1}\bar{\mathsf{g}}\rangle_{\bar{\mathsf{G}}}, \\
    \dot{\hat{\mathsfbfit{r}}}&=-\bar{\boldsymbol{\mathcal{L}}}^{\mathsf{G}}_v \hat{\mathsfbfit{r}}- \langle  \langle \bar{\mathsfbfit{N}}',\hat{\mathsfbfit{r}} \rangle_{\bar{\mathsf{E}}}, \bar{\mathsf{G}}^{-1}\bar{\mathsf{g}}\rangle_{\bar{\mathsf{G}}}, \\
    \dot{\hat{\mathsf{s}}}&=-\bar{\mathcal{L}}^{\mathsf{G}}_c \hat{\mathsf{s}}-\langle \bar{\mathsfbfit{N}}^\top, \langle  \langle \bar{\mathsfbfit{N}}',\hat{\mathsfbfit{r}} \rangle_{\bar{\mathsf{E}}} , \bar{\mathsf{G}}^{-1}\bar{\mathsf{g}}\rangle_{\bar{\mathsf{G}}} \rangle_{\bar{\mathsf{E}}}.
\end{align}
\label{eq:small1}
\end{subequations}
We can map (\ref{eq:small1}b) to (\ref{eq:small1}c) by applying $\langle \bar{\mathsfbfit{N}}^\top, \cdot \rangle_{\bar{\mathsf{E}}}$.   Likewise, $\langle \bar{\mathsfbfit{N}}, \hat{\mathsf{s}}\rangle_{\bar{\mathsf{G}}} = \bar{\boldsymbol{\mathcal{L}}}_v^{\mathsf{G}} \hat{\mathsfbfit{r}}$, so that acting on (\ref{eq:small1}c) with $\langle\bar{\mathsfbfit{N}},\cdot \rangle_{\bar{\mathsf{G}}}$ recovers $\bar{\boldsymbol{\mathcal{L}}}_v^{\mathsf{G}}$ acting on (\ref{eq:small1}b).

We assume that deformations can be captured in a basis provided by the left- and right-singular vectors, writing

\begin{equation}
    \hat{\mathsfbfit{r}}={\textstyle \sum_{p=1}^q} \alpha_p(t) \bar{\mathsfbfit{Z}}_p+{\textstyle \sum_{p=q+1}^{2N_v}} \beta_p(t) \bar{\mathsfbfit{Z}}_p, \quad     \hat{\mathsf{s}}={\textstyle \sum_{p=1}^q} \sigma_p \alpha_p(t) \bar{\mathsf{Y}}_p.
    \label{eq:pertus}
\end{equation}
The $q$ modes in which vertex motions are coupled directly to area and perimeter changes have amplitudes $\alpha_p$; remaining modes with amplitudes $\beta_p$ lie in the kernel of $\langle \bar{\mathsfbfit{N}}^\top, \cdot \rangle_{\bar{\mathsf{E}}}$, representing vertex motions that have no direct impact on area and perimeter.
Defining\begin{equation}
D_{pp'}=\left\langle \bar{\mathsfbfit{Z}}_p, \left\langle  \langle \bar{\mathsfbfit{N}}', \bar{\mathsfbfit{Z}}_{p'} \rangle_{\bar{\mathsf{E}}}, \bar{\mathsf{G}}^{-1}\bar{\mathsf{g}} \right\rangle_{\bar{\mathsf{G}}} \right\rangle_{\bar{\mathsf{E}}}
\label{eq:dpp}
\end{equation}
and noting that $D_{pp'}=D_{p'p}$,

we substitute (\ref{eq:pertus}) into (\ref{eq:small1}), project the dynamics onto indivdual modes and exploit orthogonality to recover
\begin{equation}
\left(\begin{matrix}    \dot{\alpha} \\ \dot{\beta}   \end{matrix}\right)
    =-\left( \begin{matrix}
    \mathsf{D}^{\alpha\alpha}+\mathrm{diag}(\sigma_p^2) & \mathsf{D}^{\alpha\beta} \\
    \mathsf{D}^{\alpha\beta} & \mathsf{D}^{\beta\beta}
\end{matrix}\right)
\left(\begin{matrix} {\alpha} \\ {\beta} \end{matrix}\right),
    \label{eq:linearevol}
\end{equation}
where $(\alpha,\beta)=(\alpha_1,\dots,\alpha_q,\beta_{q_+1},\dots,\beta_{N_v})$, $D^{\alpha\alpha}_{pp'}=D_{pp'}$ for $1\leq p, p'\leq q$,  $D^{\alpha\beta}_{pp'}=D_{pp'}$ for $1\leq p \leq q<p'<2N_v$ and 
$D^{\beta\beta}_{pp'}=D_{pp'}$ for $q\leq p, p'<N_v$.  Eigenvalues $\lambda^{(s)}$ of the matrix in (\ref{eq:linearevol})  ($\tilde{\mathsf{D}}$, say) recover the full spectrum of relaxation rates of the vertices.  Orthonormal eigenvectors $(\alpha^{(s)},\beta^{(s)})^\top$, with (\ref{eq:pertus}), recover the associated spatial modes $\hat{\mathsfbfit{r}}^{(s)}e^{-\lambda^{(s)}t}$, which are in turn orthonormal under $\langle \cdot,\cdot\rangle_{\bar{\mathsf{E}}}$: specifically, using (\ref{eq:svd}),
\begin{multline}
    \langle \hat{\mathsfbfit{r}}^{(s)},\hat{\mathsfbfit{r}}^{(s')}
    \rangle_{\bar{\mathsf{E}}}=
    \langle {\textstyle{\sum_{p}}}\alpha_{p}^{(s)} \mathsfbfit{Z}_{p}+ {\textstyle{\sum_p}}\beta_{p}^{(s)} \mathsfbfit{Z}_{p},{\textstyle{\sum_{p'}}}\alpha_{p'}^{(s')} \mathsfbfit{Z}_{p'}+ {\textstyle{\sum_{p'}}}\beta_{p'}^{(s')} \mathsfbfit{Z}_{p'}
    \rangle_{\bar{\mathsf{E}}}\\
    ={\textstyle{\sum_p}}\alpha_{p}^{(s)}\alpha_{p}^{(s')}+{\textstyle{\sum_p}}\beta_{p}^{(s)}\beta_{p}^{(s')}=\delta_{ss'}.
\end{multline}
Inserting $\hat{\mathsfbfit{r}}^{(s)}e^{-\lambda^{(s)}t}$ into (\ref{eq:small1}b) and then contracting with $\langle \hat{\mathsfbfit{r}}^{(s)}, \cdot \rangle_{\bar{\mathsf{E}}}$ gives
\begin{equation}
    \lambda^{(s)}={\textstyle{\sum_p}} \sigma_p^2 \left[\alpha_p^{(s)} \right]^2+ 
    {\textstyle{\sum_{p,p'}}}\alpha_p^{(s)} D_{pp'} \alpha_{p'}^{(s)} + {\textstyle{\sum_{p,p'}}}\beta_p^{(s)} D_{pp'} \beta_{p'}^{(s)}.
    \label{eq:buildspectrum}
\end{equation}
Eq.~(\ref{eq:buildspectrum}) allows us to evaluate how the decay rate of a given mode is built from eigenvalues of the Laplacian operators (the terms involving $\sigma_p^2$) and from interaction with the pre-stressed equilibrium state (the terms involving $\mathsf{D}$, which capture evolution of the operators).  The area change associated with eigenvector $\hat{\mathsfbfit{r}}^{(s)}$ is specified by non-zero components in $\alpha^{(s)}$; the associated first-order energy change $\hat{\mathsf{s}}^\top \bar{\mathsf{g}}$ is proportional to

${\textstyle \sum_{p'=1}^q} \sigma_{p'} \alpha_{p'}^{(s)} \langle \bar{\mathsf{Y}}_{p'}, \bar{\mathsf{G}}^{-1}\bar{\mathsf{g}}\rangle_{\bar{\mathsf{G}}}$; $\lambda^{(s)}$ in (\ref{eq:buildspectrum}) gives the second variation of the energy in (\ref{eq:energyvar}) for mode $s$.

In summary, reformulation of the problem using SVD has revealed the equilibrium Laplacian operators $\bar{\mathcal{L}}_c^{\mathsf{G}}$ and $\bar{\boldsymbol{\mathcal{L}}}_v^{\mathsf{G}}$ that drive the dynamics, while providing spatial bases that connect vertex dynamics to patterns of cell area and perimeter change.  Using (\ref{eq:buildspectrum}), we can now assess the relative contributions of material and geometric stiffness to the dynamics, with the former driven by Laplacian operators and the latter by the coupling 
between vertex motions and the equilibrium prestress, which perturbs these operators.  

\section{Results}
\label{sec:results}

\begin{figure}
\centering
\includegraphics[width=\textwidth]{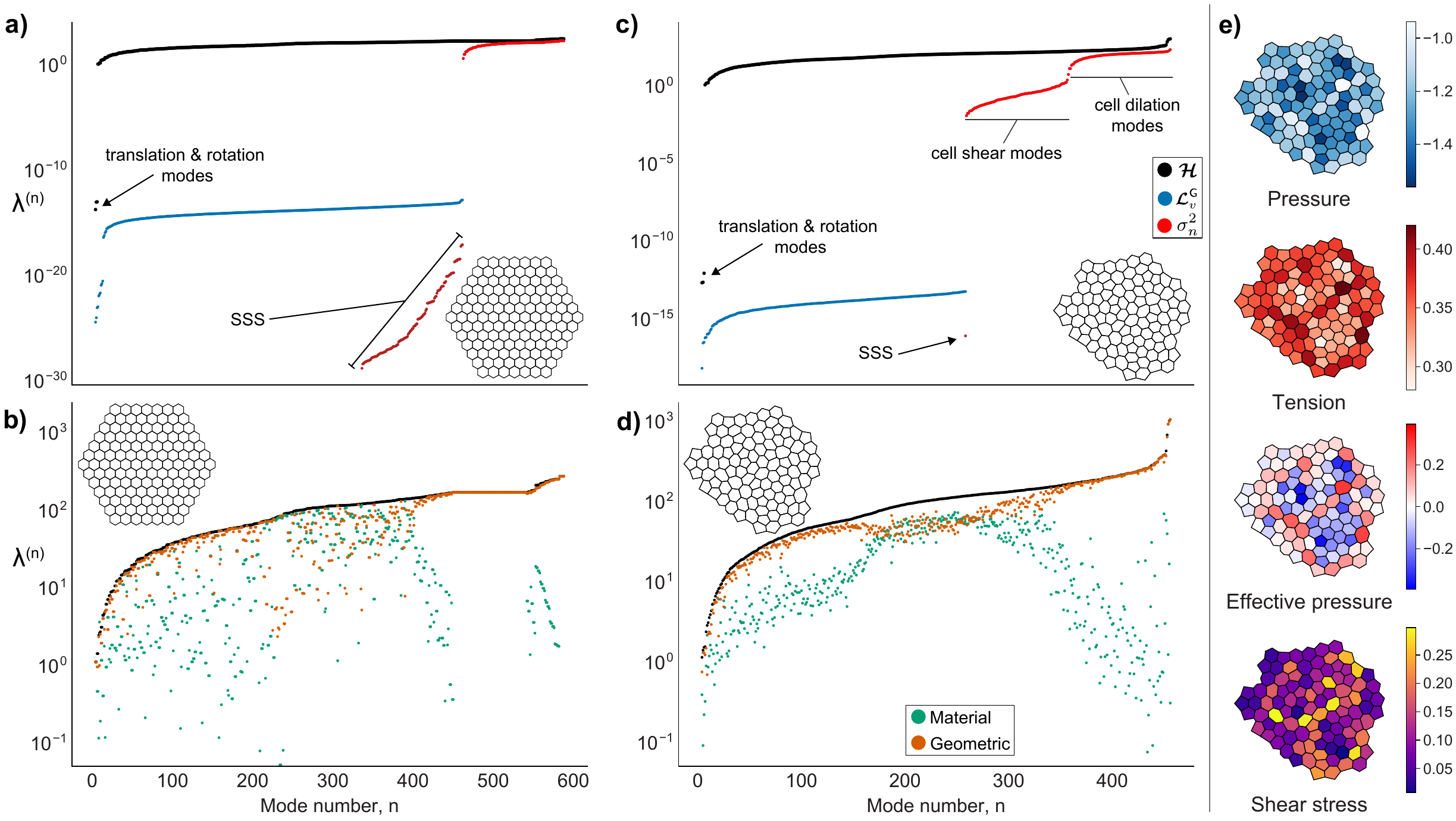}
    \caption{(a,b) Spectra for a symmetric configuration of $N_c=127$ hexagonal cells, as depicted in the inset; (c,d) spectra for a disordered monolayer of $N_c=100$ cells; $\Gamma=0.5$, $L_0=1$, $\gamma_A=\gamma_L=0$ in both cases.  (e) shows the corresponding pressure, tension and isotropic and deviatoric cell stress components $P_{\mathrm{eff},i}$ and $\zeta_i$ in (\ref{eq:peffshear}) of the equilibrium disordered monolayer. 
    (a,c) compare the eigenvalues $\lambda^{(n)}$, $n=1,\dots, 2N_v$ of the Hessian operator (\ref{eq:lin}a, black) to the $2N_v$ eigenvalues of the vertex Laplacian (\ref{eq:lapcv}, blue) and the $q$ non-zero eigenvalues of the cell Laplacian (\ref{eq:lapcv}, red); $N_v=294$ and $q=N_c$ in (a) and $N_v=228$ and $q=2N_c-1=199$ in (c); zero modes of $\bar{\boldsymbol{\mathcal{L}}}_v^\mathsf{G}$ have $\lambda^{(n)}<10^{-10}$; 
    translation and rotation modes of the Hessian are indicated; the $N_c$ zero modes in (a) and single zero mode in (c) of $\bar{\mathcal{L}}_c^{\mathsf{G}}$ are indicated as states of self-stress (SSS); cell shear and cell dilation modes are indicated in (c).
    (b, d) show the Hessian spectrum on an enlarged scale, excluding the translation and rotation zero modes, decomposed using (\ref{eq:buildspectrum}) into contributions from material stiffness (green, derived from $\mathcal{L}_c^{\mathsf{G}}$) and geometric stiffness (orange, derived from prestress).
}
\label{fig:2}
\end{figure}

Spectra for a symmetric and a disordered monolayer (each in a jammed state, with $\Gamma=0.5$, $L_0=1$) are presented in Figure~\ref{fig:2}.   The $q$ non-zero eigenvalues of the Laplacian operators $\bar{\mathcal{L}}_c^{\mathsf{G}}$ and $\bar{\boldsymbol{\mathcal{L}}}_v^{\mathsf{G}}$ overlap: $q=2N_c-1$ ($q=N_c$) for the disordered (symmetric) monolayer.  For the disordered monolayer, $\bar{\mathcal{L}}_c^{\mathsf{G}}$ therefore has a single zero eigenvalue corresponding to the state of self-stress (SSS) associated with the equilibrium, shown in Fig.~\ref{fig:2}e; $\bar{\boldsymbol{\mathcal{L}}}_v^{\mathsf{G}}$ has $2N_v-q=2N_v-2N_c+1$ zero modes (for which $\lambda<10^{-10}$ in Fig.~\ref{fig:2}c).  The symmetric monolayer possesses $2N_v-N_c$ zero modes of $\bar{\boldsymbol{\mathcal{L}}}_v^{\mathsf{G}}$ and $N_c$ zero modes of $\bar{\mathcal{L}}_c^{\mathsf{G}}$ (Fig.~\ref{fig:2}a).  We describe the two branches of $\bar{\mathcal{L}}_c^{\mathsf{G}}$ evident in Fig.~\ref{fig:2}(c) as cell shear and cell dilation modes, for reasons given below; for the symmetric monolayer, the cell shear modes have (effectively) zero eigenvalue, becoming SSS modes.  However the full spectrum of the Hessian (computed using (\ref{eq:lin}a) and via projection of the dynamics onto right-singular vectors (\ref{eq:buildspectrum})) hides this structure, with geometric stiffness dominating the bulk of the spectrum (Fig.~\ref{fig:2}b,d).  This stiffness is provided by the prestress in the monolayer.  Heterogeneities of the isotropic and deviatoric components of cell stress (Fig.~\ref{fig:2}e) reflect heterogeneities in cell pressures and tensions.  In contrast, cells in the hexagonal array have zero cell stress, although prestress is still provided by uniform cell pressures and tensions.  Geometric stiffness exploits sums of tensions and differences of pressures across edges (Appendix~\ref{sec:prst}), suggesting that tensions are the dominant source of geometric stiffness in the symmetric case.  In both examples, the Hessian has three zero modes (representing two translations and a rotation; Fig.~\ref{fig:2}a,c).  A plateau in the spectrum of the hexagonal array (Fig.~\ref{fig:2}b) reflects a symmetry in the associated modes (which involve motion of internal vertices; a similar plateau was reported in \cite{tong2023}). 

The relative contributions of material and geometric stiffness were reported for the vertex model in  \cite{damavandi2022B} in terms of the density of states of the Hessian and its two component matrices (see (\ref{eq:lin}a)), averaged over realisations, assuming periodic boundary conditions.  Our results (Fig.~\ref{fig:2}b,d) are consistent with these findings, but offer a more finely resolved picture for isolated monolayers, using instead the decomposition (\ref{eq:buildspectrum}). 

\begin{figure}
\centering
    \includegraphics[width=1\textwidth]{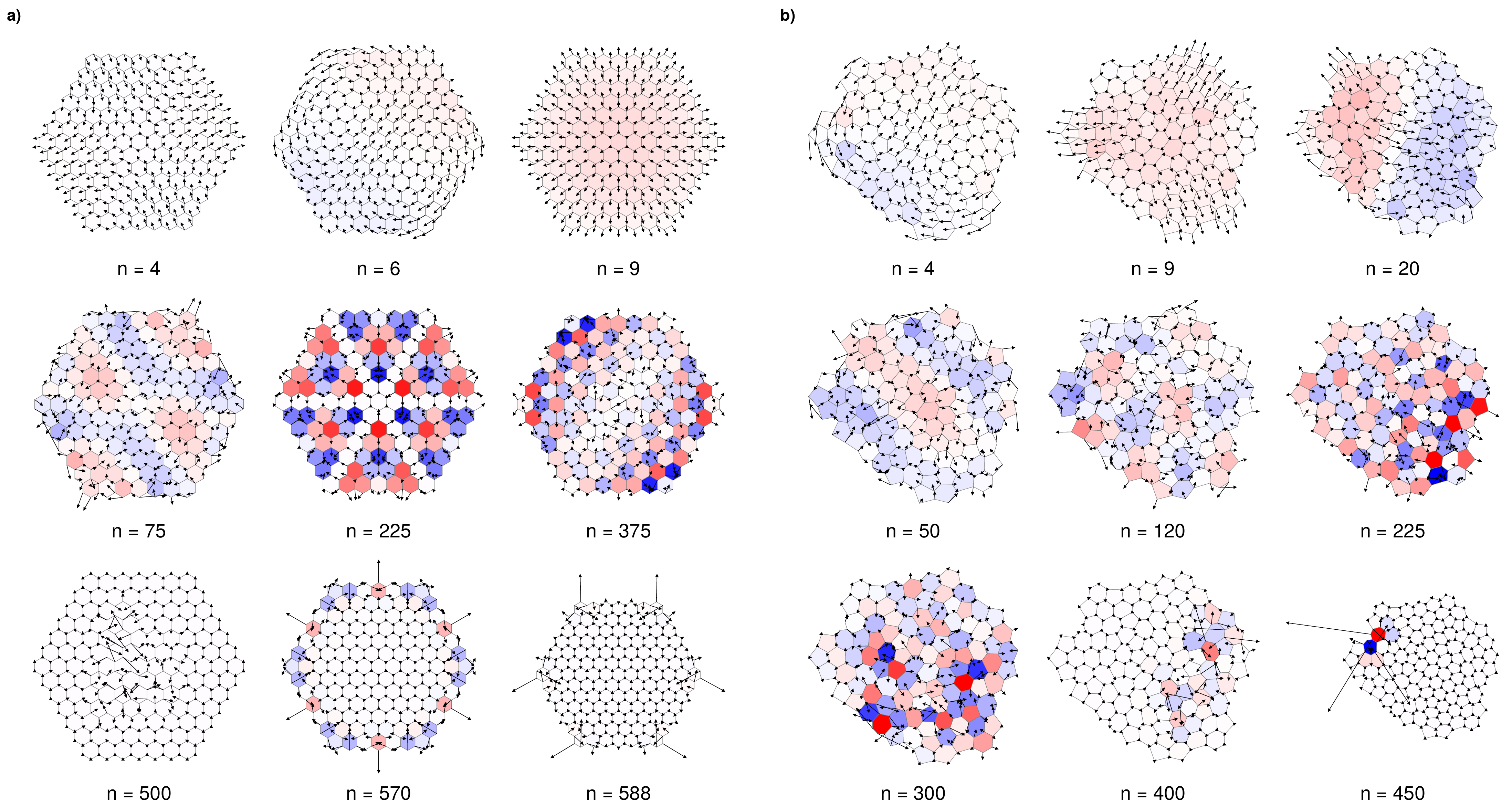}
    \caption{A selection of eigenmodes of the Hessian of (a) the hexagonal monolayer and (b) the disordered monolayer shown in Fig.~\ref{fig:2}.   The corresponding eigenvalues are ranked by index $n$; arrows show vertex displacements, colours show corresponding cell area variations, with red and blue indicating variations of opposite signs.}
    \label{fig:4}
\end{figure}

Eigenmodes of the Hessians (Fig.~\ref{fig:4}) of the disordered and symmetric monolayers involve vertex displacements that may lead to changes in cell area and perimeter; displacements are typically non-affine.  Moving along the eigenvalue spectrum, the slowest-decaying modes involve deformations over lengthscales comparable to the size of the monolayer; modes with shorter lengthscales decay more quickly.  For the symmetric monolayer, mode 500 (of 592) illustrates an internal deformation appearing in the plateau of the spectrum shown in Fig.~\ref{fig:2}(a), for which there is no change in cell area; modes 570 and 588 illustrate how the upturn at the right-hand end of the spectrum is associated with modes confined to external cells.  For the disordered monolayer, rapidly decaying modes (such as 400 and 450, of 456) are spatially localised.  

\begin{figure}
\centering
    \includegraphics[width=0.75\textwidth]{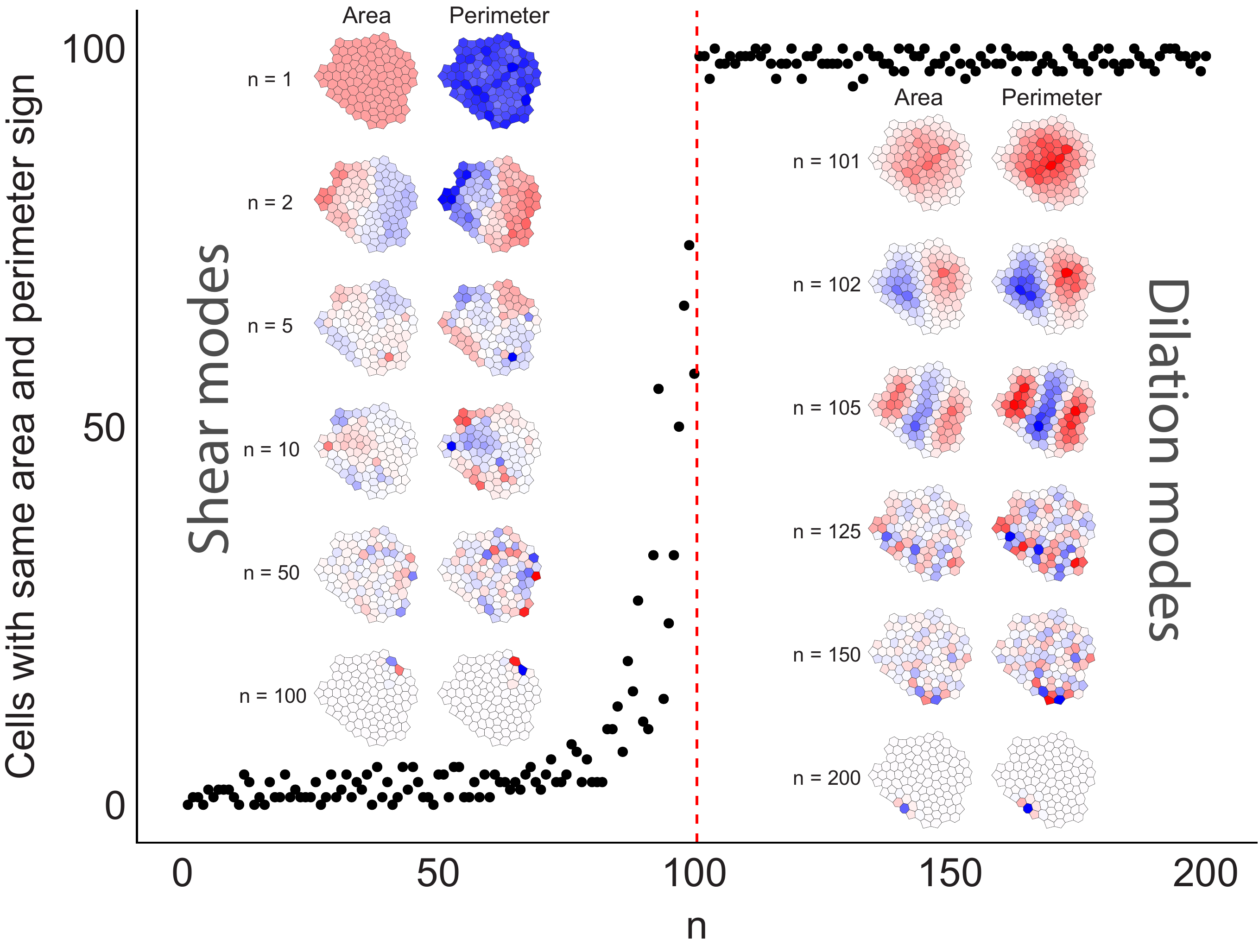}
    \caption{Insets show eigenmodes of the cell Laplacian $\bar{\mathcal{L}}_c^{\mathsf{G}}$, for the 100-cell disordered monolayer shown in Fig.~\ref{fig:2}, with mode numbers as indicated; colours indicate patterns of relative area and perimeter variation (red and blue show opposite signs).  Shear [dilation] modes (left [right]) are defined to have area and perimeter values of opposite [the same] parity; the categorisation is clear for the majority of modes, as demonstrated in the scatterplot which counts the number of cells which have the same sign for their area and perimeter values.  Mode $n=1$ is the SSS.
}
\label{fig:3}
\end{figure}

The spatial structure of the eigenmodes of the underlying scalar Laplacian $\bar{\mathcal{L}}_c^{\mathsf{G}}$ for the disordered monolayer are illustrated in Fig.~\ref{fig:3}.  These form two distinct groups, representing cell shear and cell dilation.  Dilation modes relax faster than shear modes (Fig.~\ref{fig:2}c), with pressure and tension perturbations having the same sign; shear modes typically have pressure and tension perturbations of opposite sign (Fig.~\ref{fig:3}).  Labelling these modes with $1\leq n\leq 2N_c$, mode $n=1$, with eigenvalue zero (the SSS), has $\mathsf{G}\mathsf{Y}_1$ proportional to $\overline{\mathsf{g}}$ (see (\ref{eq:bars})).  Low-order shear modes (\hbox{e.g.} $n=2,5$) have lengthscales comparable to the whole monolayer, with area and perimeter perturbations of opposite sign.  As the mode number increases towards $N_c$, the eigenmodes become increasingly localised.  Long-wave cell dilation modes appear for modes $N_c+1=101$, $N_c+2=102$, this time with area and perimeter perturbations having the same sign.  Again, these modes become increasingly localised as the mode number increases towards $2N_c$.  The corresponding eigenmodes of $\bar{\boldsymbol{\mathcal{L}}}_v^{\mathsf{G}}$ (giving more fine-grained vertex displacements) are readily determined via (\ref{eq:zeromodes}a).

\begin{figure}
\centering
    \includegraphics[width=1\textwidth]{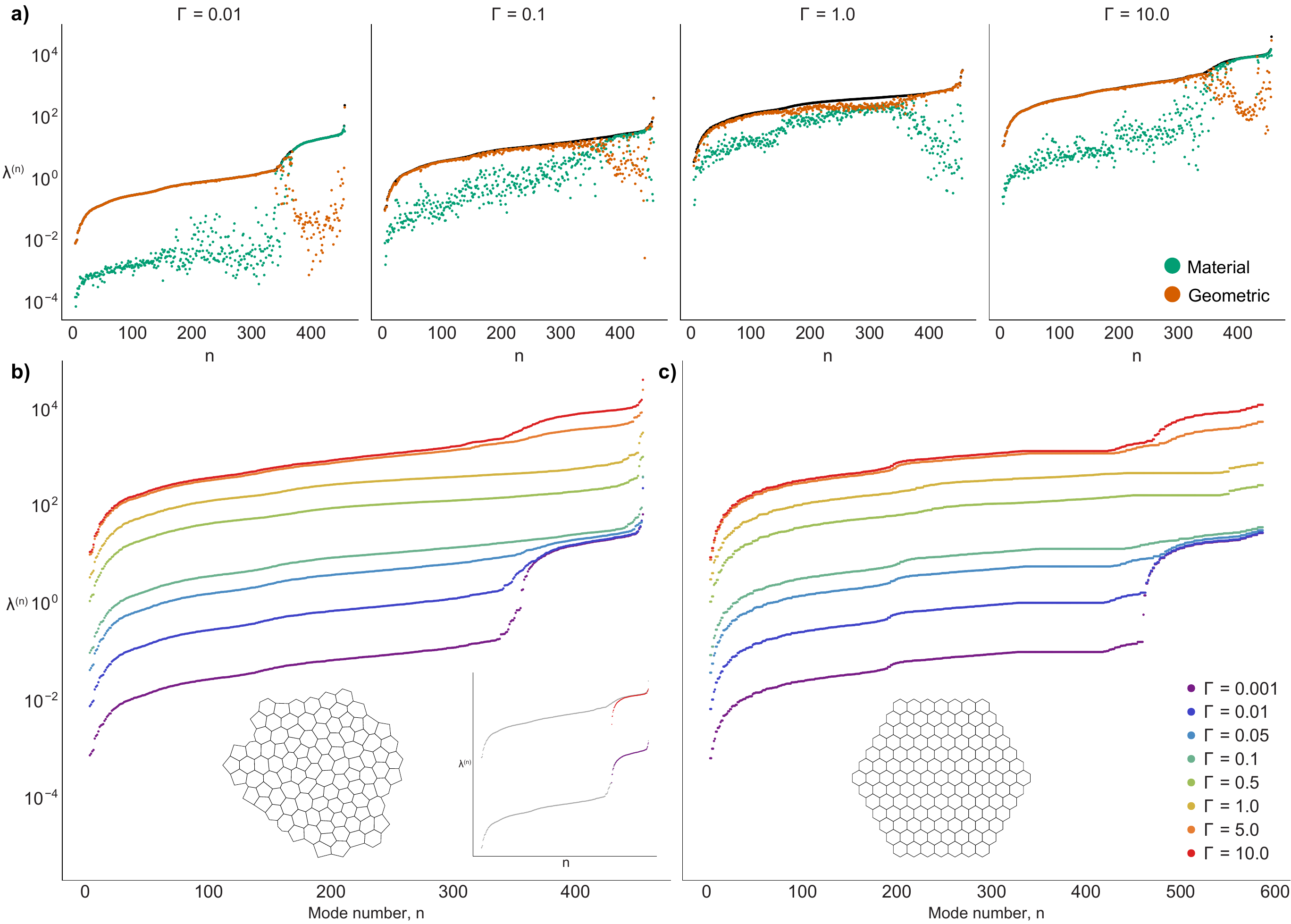}
    \caption{(a) Spectra for $L_0=1$ and $\Gamma=0.01, 0.1, 1, 10$ for realisations of disordered monolayers with $N_c=100, N_v=228$, showing contributions of geometric (orange) and material (green) stiffness to the full Hessian (black; translation and rotation modes are not shown). (b) Cell dilation modes collapse for small $\Gamma$ (blue, purple); the remainder of the spectrum collapses for large $\Gamma$ (red, orange). The inset shows how the spectra of $\mathsf{A}_c^{-1}\mathcal{L}_A$ for $\Gamma=0.001$ (purple) and $\Gamma L_0 \mathsf{L}_{c}^{-1} \mathcal{L}_{L}$ for $\Gamma=10$ (red) predict the most rapidly decaying component of the full spectrum (grey). (c) The same as (b) but for a monolayer of hexagons with $N_c=127, N_v=294$.}
    \label{fig:6}
\end{figure}

\begin{figure}
\centering
    \includegraphics[width=1\textwidth]{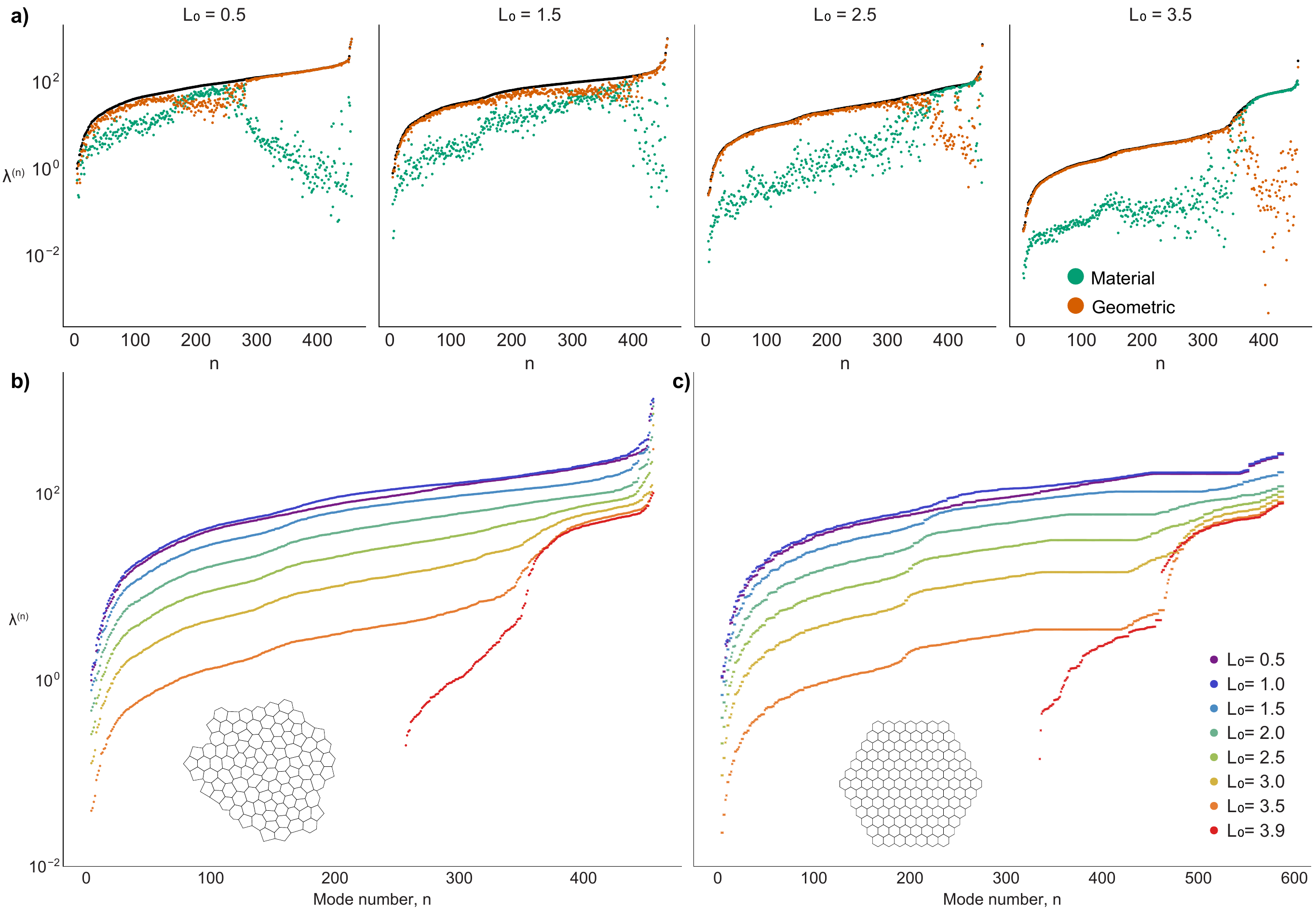}
    \caption{(a) Spectra for $\Gamma=0.5$ and $L_0=0.5,1.5, 2.5, 3.5$ for realisations of disordered monolayers with $N_c=100, N_v=228$, showing contributions of geometric (orange) and material (green) stiffness (translation and rotation modes are not shown). The case $L_0=0.5$, $\Gamma=1$ is shown in Fig.~\ref{fig:2}(d).  (b) Cell dilation modes collapse (orange, red) as $L_{0}$ approaches the critical value for the rigidity transition. After the rigidity transition (red) the Hessian spectrum completely collapses to the cell Laplacian spectrum and has $2 N_c$ non-zero eigenmodes.(c) The same as (b) but for a monolayer of hexagons with $N_c=127, N_v=294$.}
    \label{fig:7}
\end{figure}

It is known \cite{damavandi2022B} that near the jamming transition, $\bar{\mathcal{L}}_c^{\mathsf{G}}$ (in our notation) may directly determine some modes of the Hessian.  To establish more generally when this may arise, we computed spectra for different values of $\Gamma$ and $L_0$ (Figs~\ref{fig:6}, \ref{fig:7}).  For extreme values of $\Gamma$, the $N_c$ cell dilation modes capture the most rapidly decaying component of the spectrum (Fig.~\ref{fig:6}a), indicating that this component of the dynamics can be described by (\ref{eq:dilation}a) (with $\gamma_A=0$ and $\Gamma\ll 1$) or (\ref{eq:dilation}b) (with $\gamma_L=0$ and $\Gamma \gg 1$).  This is further demonstrated by the inset to Fig.~\ref{fig:6}(b), which shows how the most rapidly decaying modes of the full spectrum are resolved by the discrete approximation of $\nabla^2$ (Appendix~\ref{app:hh}) when $\Gamma$ is small. (In contrast, cell shear modes do not appear to play such a dominant role in determining the Hessian.)   
Thus, for $\Gamma=10$ (Fig.~\ref{fig:6}a), tension dominates pressure and cells are close to their target perimeter: dilation modes determine the fastest segment of spectrum via (\ref{eq:dilation}b) with eigenvalues scaling with $\Gamma$; remaining modes are dominated by geometric stiffness, but show insenstivity to $\Gamma$ (Fig.~\ref{fig:6}b).  For $\Gamma=0.01$ (Fig.~\ref{fig:6}a), pressure dominates tension and cells are close to their target area: cell dilation modes determine the fastest segment of spectrum via (\ref{eq:dilation}a), with eigenvalues of order unity (showing $\Gamma$-independence in Fig.~\ref{fig:6}b); remaining modes are dominated by geometric stiffness, scaling with (small) $\Gamma$.  For intermediate $\Gamma$ (\hbox{e.g.} $\Gamma=1$), with a balance between tension and pressure, geometric stiffness influences the entire spectrum.  A similar picture emerges for the symmetric monolayer (Fig.~\ref{fig:6}c).  In this instance, we find that the blocks of $\mathcal{L}$ in (\ref{eq:sprob}) satisfy $\mathcal{L}_L=\mathcal{L}_C^\top \mathcal{L}_A^{-1}\mathcal{L}_C$ and $\mathcal{L}_A=\mathcal{L}_C\mathcal{L}_L^{-1}\mathcal{L}_C^\top$, ensuring that shear modes have zero eigenvalue (Appendix~\ref{app:gam}).

Increasing $L_0$ towards a critical value reduces cell tensions towards zero and takes the jammed monolayer to an unjammed state.  The approach to this transition is illustrated in Fig.~\ref{fig:7}(a).  At $L_0=3.5$, the dilation modes are evident, but the eigenvalues of the remaining modes (deriving stiffness from prestress, particularly via tensions) drop substantially in magnitude, with only (slow) shear modes remaining once $L_0=3.9$ (Fig.~\ref{fig:7}b).  This is indicative of a switch from a jammed to an unjammed state.  The transition takes place at values of $L_0$ broadly consistent with prior studies using similar constitutive models (\hbox{e.g.} \cite{tong2023}).  A similar loss of geometric stiffness is evident for the symmetric monolayer (Fig.~\ref{fig:7}c).

\begin{figure}
\begin{center}

\includegraphics[width=1\textwidth]{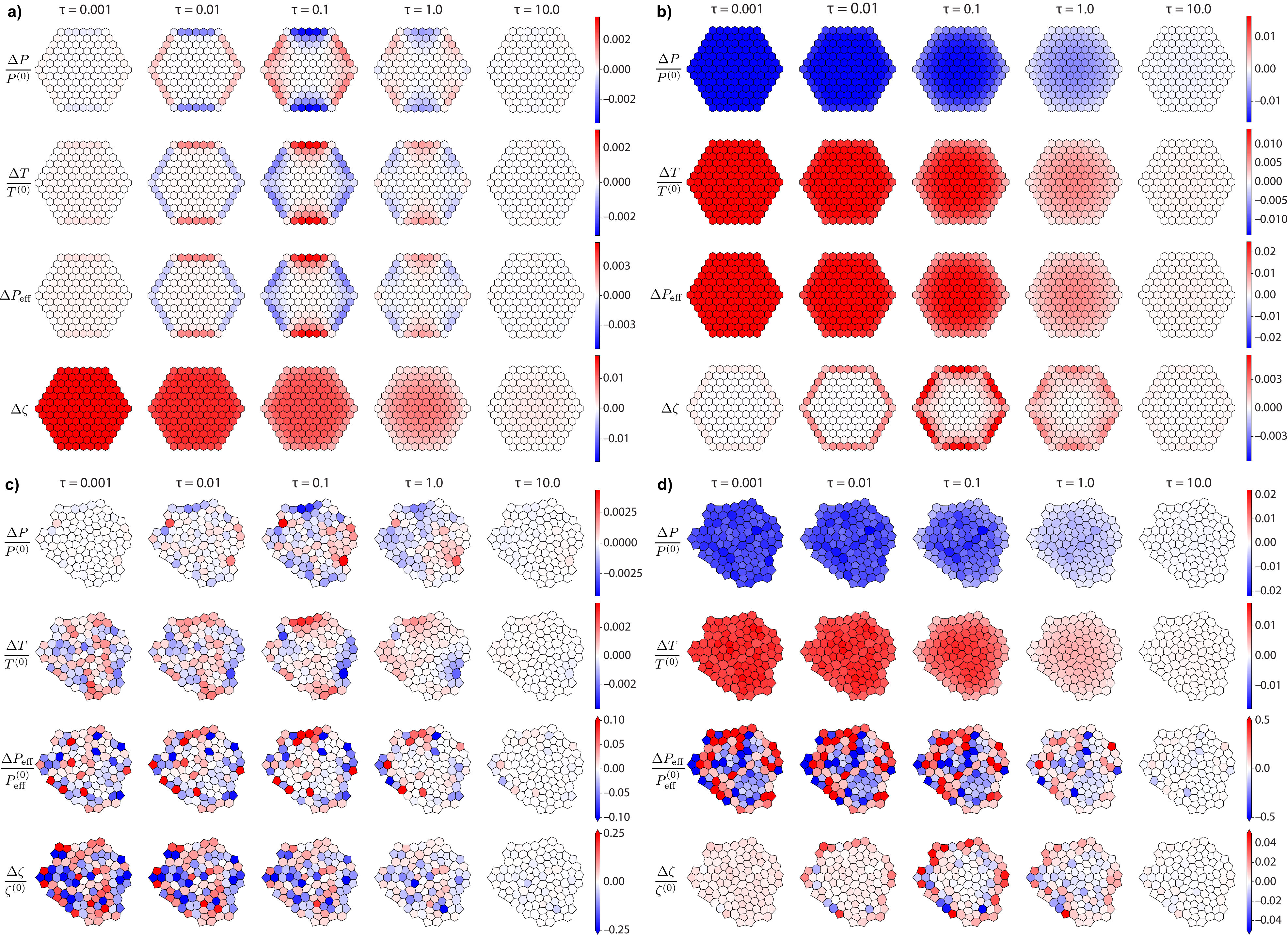}
\end{center}
\caption{ 
Changes in cell pressure, tension, isotropic stress $P_{\mathrm{eff}}$ and shear stress $\zeta$ (\ref{eq:peffshear}) immediately after 1\% uniaxial (a,c) and biaxial (b,d) stretch imposed over five timescales $\tau$ to the hexagonal (a,b) and disordered (c,d) monolayers shown in Fig.~\ref{fig:2}(c,d).   Most cases show relative changes (\hbox{i.e.} relative changes in magnitude) using normalisations as indicated; absolute changes in $P_{\mathrm{eff}}$ and $\zeta$ are reported for hexagonal cells.  Since $P_i<0$ ($T_i>0$) because $A_i<1$ ($L_i>L_0$), an increase in area (perimeter) under stretch leads to a reduction (increase) in the magnitude of $P_i$ ($T_i$).  Subsequent relaxation of pressure and tension for $\tau=0.1$ is illustrated for disordered cells in Supplementary Movies 1 (uniaxial) and 2 (biaxial).}
\label{fig:stretch}
\end{figure}

Returning to the baseline hexagonal and disordered monolayers illustrated in Fig.~\ref{fig:2}, we imposed a 1\% stretch on each monolayer over a time interval $\tau$, implemented as described in Appendix~\ref{app:stretch}.  Stretch is modelled via a viscous drag imposed on vertices; the amplitude is too small to induce neighbour exchanges, and the monolayer relaxes to its original equilibrium after the stretch terminates.  The five chosen rates $1/\tau$ span the eigenvalue spectra.  Considering first the response of the hexagonal monolayer (Fig.~\ref{fig:stretch}a), fast uniaxial stretch generates a shear stress response that is distributed uniformly across the monolayer, becoming weaker for slower stretch.  At intermediate stretch rates however, peripheral cells show a different response to those in the bulk: the shear stress response is weakened, with peripheral cells along the top and bottom boundaries (aligned with the main axis of stretch) showing evidence of expansion and those along the remaining boundaries showing evidence of compression.  In contrast, fast biaxial loading of the hexagonal monolayer (Fig.~\ref{fig:stretch}b) generates uniform cell expansion, as reflected in the isotropic stress $P_{\mathrm{eff}}$; however at intermediate loading rates, peripheral cells experience shear stress that appears to be associated with a reduction in the magnitude of the $P_{\mathrm{eff}}$ response.  

Similar features are seen in the disordered monolayer, although these are modified by its intrinsic heterogeneity.  The shear stress response to rapid uniaxial stretching (Fig.~\ref{fig:stretch}c) is distributed across the monolayer; slower stretching weakens this response but induces expansion and compression of peripheral cells, although both patterns are heterogeneous.  Rapid biaxial stretching (Fig.~\ref{fig:stretch}d) generates uniform cell expansion but no appreciable shear stress; again, at intermediate stretching rates, the changes to pressures and tensions in peripheral cells are weaker than those in the interior, although the peripheral cells now experience some shear stress. Supplementary Movies 1 and 2 show how pressures and tensions relax after unaxial and biaxial stretch; in the latter case, a largely homogeneous disturbance relaxes quickly, leaving slower evolution of a more heterogeneous pattern.   In both examples, very slow loading ($\tau=10$) is too weak to elicit a strong response.  In summary, Fig.~\ref{fig:stretch} shows how patterns of geometric changes (reflected in pressures and tensions) can differ significant from patterns of stress changes, showing sensitivity to the mode and rate of deformation; furthermore, cells at the monolayer periphery can experience deformations and stresses that are distinct from those in the interior.

\section{Discussion}
\label{sec:disc}

Despite being relatively straightforward to formulate and to simulate, the cell vertex model exemplifies many of the outstanding challenges in multiscale modelling, particularly concerning the relationship between microscopic and macroscopic mechanics of cellular materials.  Here we have used spectral techniques to investigate the relaxation of a multicellular material to an equilibrium state, as may arise after an imposed perturbation to a biological tissue.  This approach is valuable in capturing deformations across all lengthscales, avoiding any assumptions of spatial  or statistical regularity.  Even in the presence of symmetry, the dynamics can be quite distinct from that expected from conventional continuum models.

We introduced three variations to the standard vertex model, by (i) modifying dissipation to account for the size of the region adjacent to each cell vertex undergoing drag as it slides over a substrate, (ii) incorporating viscous resistance to changes in cell area and perimeter and (iii) introducing a mechanical energy that incorporates nonlinear pressure-area and tension-length relations, relaxing the requirement implicit in some conventional models that cell areas and perimeters should be everywhere close to preferred values.  These three features influence metric matrices $\mathsf{E}$, $\mathsf{H}$ and $\mathsf{G}$ respectively that appear in discrete differential operators within the model, and therefore influence stress relaxation.  While most of our qualitative observations are insensitive to detailed modelling assumptions, (i) and (iii) were chosen specifically to highlight how the operator driving cell dilation modes when cell bulk stiffness dominates cortical stiffness ($-\mathsf{A}_c^{-1}\mathcal{L}_A$ in (\ref{eq:dilation}a)) approximates the conventional spatial operator $\nabla^2$ (Appendix~\ref{app:hh}).  The operator governing cell dilation modes when cortical stiffness dominates bulk stiffness ($-\mathsf{L}_c^{-1}\mathcal{L}_L$) is a discrete spatial second derivative with a more nonlocal structure.

Using the monolayer as a network on which discrete differential operators can be constructed, we have clarified the relationship between the gradient-flow structure of the vertex model and generalised gradient and divergence operators (Figure~\ref{fig:svdsummary}) that are used to build Laplacian operators.  While these encompass the operators $-A_c^{-1}\mathcal{L}_A$ and $-\mathsf{L}_c^{-1}\mathcal{L}_L$ mentioned above, the generalised operators built from the matrix $\mathsfbfit{M}$ and its transpose have a more abstract structure, mapping between vertex displacements (in $\mathbb{R}^2$) and tangent spaces $\mathcal{TY}$ and $\mathcal{TX}$ that hold pairs of scalar attributes per cell (respectively area \& perimeter variations, and pressure \& tension variations).  Because $\mathsfbfit{M}\cdot$ and $\mathsfbfit{M}^\top$ are singular (to accommodate an equilibrium SSS, and zero modes for which vertex motions do not change areas and perimeters), SVD provides a natural framework with which to formally establish properties of the Laplacian operators $\mathcal{L}_c^{\mathsf{G}}$ and $\boldsymbol{\mathcal{L}}_v^{\mathsf{G}}$ and their duals $\mathcal{L}_c^{\mathsf{G}\dag}$ and $\boldsymbol{\mathcal{L}}_v^{\mathsf{G}\dag}$, which underpin the model; their actions are illustrated in Fig.~\ref{fig:svdsummary}(a).

Despite suggesting predominantly diffusive dynamics, the Laplacian operators evolve as the monolayer deforms, to the extent that geometric stiffness typically dominates material stiffness (Figs~\ref{fig:2}, \ref{fig:6}, \ref{fig:7}).  Prestress in an equilibrium state can therefore have a important influence on the rate at which disturbances decay.  A consistent exception is the manner in which the cell dilation modes of ${\mathcal{L}}_c^{\mathsf{G}}$ capture the fastest $N_c$ modes of the Hessian for extreme values of $\Gamma$ (Fig.~\ref{fig:6}).

While our primary interest is in monolayers in a jammed state, the model illustrates how rigidity is lost through an increase in $L_0$ (Fig.~\ref{fig:7}) via the well-studied jamming/unjamming transition \cite{bi2015}, with cell dilation modes (which exhibit resistance to area change, but not to shear) contributing primarily to monolayer stiffness past the unjamming transition.  Rigidity can also be lost through reduction in $\Gamma$ towards zero (Fig.~\ref{fig:6}).  Identification of the full spectrum of relaxation times enables us to assess the impact of transient stretch on an isolated monolayer in the jammed state (Fig.~\ref{fig:stretch}).  Noting that force regimes acting on epithelial tissue \textit{in vivo} can vary widely, it is instructive to consider differences between fast and slow rates of stretch in terms of cell shape changes (leading to changes in cell pressure and tension) and cell stress (changing the isotropic cell stress $P_{\mathrm{eff},i}$ and the cell shear stress $\zeta_i$).  Our results reveal a heterogeneous response to homogeneous loading at intermediate loading rates, with peripheral cells showing the greatest relative change in tension under uniaxial loading and in shear stress under biaxial loading (Fig.~\ref{fig:6}).  This distinct peripheral response is lost under fast stretch, and the response as a whole weakens in magnitude under slow stretch. In the future, it will be of interest to investigate how these findings relate to the biological mechano-responses (e.g. cell division) of epithelial tissue to stretch and whether these responses vary according to loading rate.

The vertex model is a gradient flow (Appendix~\ref{app:grfl}), but not (typically) in a Euclidean metric.

Nevertheless, motivated by \cite{natale2023}, it is instructive to compare the present model with a continuous gradient-flow model for a distribution function $\rho(\mathbf{x},t)$ identifying particle locations in $\mathbb{R}^2$.  For example, noting that cell areas are quadratic and symmetric in $\mathsfbfit{r}$ (and ignoring tension effects temporarily), a kernel $K(\mathbf{x};\mathbf{y},\mathbf{z},t)$ can be used to construct an area-like variable $\mathcal{A}(\mathbf{x},t)=\tfrac{1}{2}\iint K(\mathbf{x};\mathbf{y}, \mathbf{z},t)\rho(\mathbf{y},t)\rho(\mathbf{z},t)\,\mathrm{d}\mathbf{y}\,\mathrm{d}\mathbf{z}$.  The dependence of $K$ on $\mathbf{x}$ encodes the (potentially evolving) cellular microstructure and we impose $K(\mathbf{x};\mathbf{y},\mathbf{z},t)=K(\mathbf{x};\mathbf{z},\mathbf{y},t)$.  Using a suitable convex functional $\mathcal{U}[\mathcal{A}]$, which defines a pressure $\mathcal{P}(\mathbf{x},t)=\delta\mathcal{U}/\delta\mathcal{A}$, one can construct an energy $U(t)=\int \mathcal{U}[\mathcal{A}(\mathbf{x},t)]\,\mathrm{d}\mathbf{x}$, such that \begin{equation}
    \frac{\delta U}{\delta \rho}(\mathbf{x};\mathbf{y},t)=\mathcal{P}(\mathbf{x},t)\frac{\delta A}{\delta \rho}(\mathbf{x};\mathbf{y},t), \quad  \frac{\delta A}{\delta \rho}(\mathbf{x};\mathbf{y},t)=\int K(\mathbf{x};\mathbf{y},\mathbf{z},t)\rho(\mathbf{z},t)\,\mathrm{d}\mathbf{z}.
\end{equation}
Then, for some positive drag functional $E[\rho]>0$, kinematics and the analogue of (\ref{eq:v}) give 
\begin{equation}
\rho_t+\nabla\cdot (\rho \mathbf{v})=0, \quad E[\rho(\mathbf{y},t)]\mathbf{v}(\mathbf{y},t)=-\nabla_{\mathbf{y}} \int \frac{\delta U}{\delta\rho}(\mathbf{x}; \mathbf{y},t)\,\mathrm{d}\mathbf{x}, \quad 
\label{eq:cont}
\end{equation}
so that under suitable boundary conditions,
\begin{align}
U_t&=\iint \mathcal{P}(\mathbf{x},t) \frac{\delta A}{\delta\rho}(\mathbf{x};\mathbf{y},t)\rho_t(\mathbf{y},t)\,\mathrm{d}\mathbf{x}\,\mathrm{d}\mathbf{y}
=-\iint \mathcal{P}(\mathbf{x},t) \frac{\delta A}{\delta\rho}(\mathbf{x};\mathbf{y},t)\nabla_\mathbf{y}\cdot [\rho(\mathbf{y},t)\mathbf{v}(\mathbf{y},t)]\,\mathrm{d}\mathbf{x}\,\mathrm{d}\mathbf{y}\\
&=\iint \rho(\mathbf{y},t)\mathbf{v}(\mathbf{y},t)\cdot \nabla_\mathbf{y} \left[ \mathcal{P}(\mathbf{x},t) \frac{\delta A}{\delta\rho}(\mathbf{x},\mathbf{y},t)\right]\,\mathrm{d}\mathbf{x}\,\mathrm{d}\mathbf{y}=-\int \rho E[\rho] \mathbf{v}\cdot \mathbf{v} \,\mathrm{d}\mathbf{y}\leq 0,
\end{align}
which recalls (\ref{eq:diss}).  The evolution equation (\ref{eq:cont}) reduces to the system 

\begin{subequations}
\label{eq:enden}
\begin{align}
    \rho_t(\mathbf{y},t)&=\int \mathcal{P}(\mathbf{x},t) \nabla_{\mathbf{y}}\cdot \left(\frac{\rho(\mathbf{y},t)} {E[\rho(\mathbf{y},t)]} \int \nabla_\mathbf{y} K(\mathbf{x};\mathbf{y},\mathbf{z},t)\rho(\mathbf{z},t)\,\mathrm{d}\mathbf{z}\right)\,\mathrm{d}\mathbf{x}, \\
    \mathcal{P}(\mathbf{x},t)&=\mathcal{U}'\left[ \tfrac{1}{2}\iint K(\mathbf{x};\mathbf{y},\mathbf{z},t)\rho(\mathbf{y},t)\rho(\mathbf{z},t)\mathrm{d}\mathbf{y}\mathrm{d}\mathbf{z}\right].
\end{align}
\end{subequations}
Additional terms could be included to account for tension effects.  The resulting nonlocal nonlinear system falls into the broad category of aggregation-diffusion models \cite{carrillo2019}, and  
suggests a route to a continuum representation of force chains \cite{liu2021}, non-affine and long-range deformations \cite{bose2019, lerner2023} and related hyperbolic effects \cite{blumenfeld2004}.

In summary, we have shown how SVD provides a natural framework with which to relate cell-level and vertex-level dynamics in the relaxation of an isolated epithelial monolayer.  Despite a clear role for scalar Laplacians (identifying $2N_c$ degrees of freedom related to material stiffness, driving primarily diffusive dynamics), simulations confirm the central role of geometric stiffness mediated by $2N_v$ degrees of freedom in determining the full relaxation spectrum.  Considering the model as a gradient flow (\hbox{e.g.} (\ref{eq:enden})), we have also shown how discrete approximations of some familiar spatial operators compete with more exotic nonlocal derivatives. 

\ack{This work was supported by The Leverhulme Trust (RPG-2021-394), the Biotechnology and Biological Sciences Research Council (BB/T001984/1) and the Wellcome Trust (225408/Z/22/Z; 210062/Z/17/Z).  For the purpose of open access, the authors have applied a Creative Commons Attribution (CCBY) licence to any Author Accepted Manuscript version arising.}

%\begin{appendix}
\appendix
\numberwithin{equation}{section}

\section*{Appendix}

\section{Geometric operators $\mathsfbfit{M}$ and $\mathsfbfit{M}'$ and the shape tensor $\mathsfbfit{Q}$}
\label{sec:appx}

To calculate the geometric operators $\mathsfbfit{M}$ and $\mathsfbfit{M}'$, it is helpful to introduce signed incidence matrices $\mathsf{A}$ and $\mathsf{B}$, which capture the topology of the cell network.  We adopt notation used in  \cite{jensen2020, jensen2022}: $A_{jk}=1$ (or $-1$) if directed cell edge $j$ points into (out of) vertex $k$ and is zero otherwise; $B_{ij}=1$ (or $-1$) if directed cell edge $j$ is congruent (anticongruent) with the orientation of the face of cell $i$, and is zero otherwise.  We define $\overline{A}_{jk}\equiv \vert A_{jk}\vert$ and $\overline{B}_{ij}\equiv \vert B_{ij}\vert$.  All cell faces are assigned the same orientation $\boldsymbol{\epsilon}_i$, which is the $2\times 2$ matrix representing a $\pi/2$ rotation; cell edge orientations are assigned arbitrarily.  Edge vectors are then related to vertex locations by $\mathbf{t}_j=\sum_k A_{jk}\mathbf{r}_k$.  Three useful identities are
\begin{equation}
    \frac{\partial \mathbf{t}_j}{\partial \mathbf{r}_m}=A_{jm}\mathsf{I}_2, \quad
    \frac{\partial {t}_j}{\partial \mathbf{r}_m}=A_{jm}\check{\mathbf{t}}_j, \quad
    \frac{\partial \check{\mathbf{t}}_j}{\partial \mathbf{r}_m}=A_{jm}\frac{\mathsf{I}_2-\check{\mathbf{t}}_j\otimes \check{\mathbf{t}}_j}{t_j}.
\label{eq:3ids}
\end{equation}
Here a check denotes a unit vector, so that $\check{\mathbf{t}}_j=\mathbf{t}_j/t_j$.  We define $\mathbf{n}_{ij}=-B_{ij}\boldsymbol{\epsilon}_i \mathbf{t}_j$ to be the outward normal to cell $i$ at edge $j$ and $\mathbf{c}_j=\tfrac{1}{2}\sum_k\overline{A}_{jk}\mathbf{r}_k$ to be the centroid of edge $j$.  The link $\mathbf{s}_{ik}$ in cell $i$ between adjacent edge centroids at vertex $k$ (Fig.~\ref{fig:hh}a) and its normal $\mathbf{n}_{ik}$ pointing into the cell satisfy 
\begin{equation}
\mathbf{s}_{ik}={\textstyle{\sum_j}} \tfrac{1}{2} B_{ij} \mathbf{t}_j \overline{A}_{jk}, \quad 
\mathbf{n}_{ik}\equiv \boldsymbol{\epsilon}_i\mathbf{s}_{ik}=-\tfrac{1}{2} {\textstyle{\sum_j}} \overline{A}_{jk}\mathbf{n}_{ij}.
\label{eq:sik}
\end{equation}

Integrating over cell $i$ with area $A_i$, $\mathsf{I}_2 A_i=\int_i \nabla \otimes \mathbf{x} \,\mathrm{d}A=\oint_i \check{\mathbf{n}}\otimes \mathbf{x}\,\mathrm{d}s=\sum_j \check{\mathbf{n}}_{ij}\otimes \oint\mathbf{x}\,\mathrm{d}s=\sum_j\mathbf{n}_{ij}\otimes \mathbf{c}_j=-\sum_j B_{ij}\boldsymbol{\epsilon}_i \mathbf{t}_j \otimes \mathbf{c}_j$.  Thus small variations in area are related to small changes in vertex locations by via
\begin{equation}
\mathsf{I}_2\,{\delta A_i}={\textstyle{\sum}_{j,k}}\left\{ - \tfrac{1}{2} {\textstyle\sum_{k'}} B_{ij}A_{jk} \overline{A}_{jk'}\boldsymbol{\epsilon}_i\delta\mathbf{r}_k\otimes \mathbf{r}_{k'}-\tfrac{1}{2}  B_{ij}\overline{A}_{jk}\boldsymbol{\epsilon}_i\mathbf{t}_j \otimes \delta\mathbf{r}_k \right\}.
\end{equation}
The identity 
\begin{equation}
{\textstyle \sum_j} B_{ij}A_{jk'}\overline{A}_{jk}=-{\textstyle \sum_j} B_{ij}\overline{A}_{jk'}A_{jk}
\label{eq:BAAid}
\end{equation}
gives
\begin{align}
\mathsf{I}_2\,{\delta A_i}&={\textstyle{\sum}_{j,k}}\left\{  \tfrac{1}{2}{\textstyle{\sum_{k'}}} B_{ij}\overline{A}_{jk} {A}_{jk'}\boldsymbol{\epsilon}_i\delta\mathbf{r}_k\otimes \mathbf{r}_{k'}-\tfrac{1}{2}  B_{ij}\overline{A}_{jk}\boldsymbol{\epsilon}_i\mathbf{t}_j \otimes \delta\mathbf{r}_k \right\} \nonumber \\ &
={\textstyle{\sum}_{j,k}}  \tfrac{1}{2} B_{ij}\overline{A}_{jk} \left\{\boldsymbol{\epsilon}_i\delta\mathbf{r}_k\otimes \mathbf{t}_{j}-  \boldsymbol{\epsilon}_i\mathbf{t}_j \otimes \delta\mathbf{r}_k \right\}.
\end{align}
Taking the trace,
\begin{align}
2{\delta A_i}&={\textstyle{\sum}_j}  \tfrac{1}{2} B_{ij}\overline{A}_{jk} \left\{(\boldsymbol{\epsilon}_i \delta\mathbf{r}_k)^T \mathbf{t}_{j}- ( \boldsymbol{\epsilon}_i\mathbf{t}_j)^T \delta\mathbf{r}_k \right\}
={\textstyle{\sum}_j}  \tfrac{1}{2} B_{ij}\overline{A}_{jk} \left\{-\delta\mathbf{r}_k^T \boldsymbol{\epsilon}_i  \mathbf{t}_{j}- \delta\mathbf{r}_k^T  \boldsymbol{\epsilon}_i\mathbf{t}_j  \right\}
\end{align}
implying (using (\ref{eq:sik}))
\begin{subequations}
    \begin{equation}
\frac{\partial A_i}{\partial \mathbf{r}_k}=-\tfrac{1}{2} {\textstyle{\sum}_j}  B_{ij}\overline{A}_{jk} \boldsymbol{\epsilon}_i  \mathbf{t}_{j} \equiv \tfrac{1}{2}{\textstyle\sum_{j}}\overline{A}_{jk}\mathbf{n}_{ij}=-\mathbf{n}_{ik}.
\label{eq:Aik}
\end{equation}
Noting that $L_i=\sum_j \overline{B}_{ij}t_j$, we use (\ref{eq:3ids}) to give 
\begin{equation}
    \frac{\partial L_i}{\partial \mathbf{r}_k}={\textstyle{\sum_j}}\overline{B}_{ij} A_{jk}\check{\mathbf{t}}_j.
    \label{eq:Lik}
\end{equation}
\label{eq:ALik}
\end{subequations}
We use (\ref{eq:ALik}) to construct the elements $\mathbf{M}_{\alpha k}$ of $\mathsfbfit{M}$.  It also allows us to express the vertex evolution equation (\ref{eq:v}) using incidence matrices as
\begin{equation}
E_k \dot{\mathbf{r}}_k={\textstyle{\sum_{i,j}}} \left(\tfrac{1}{2} {P}_i B_{ij}\boldsymbol{\epsilon}_i \mathbf{t}_j \overline{A}_{jk}-{T}_i \overline{B}_{ij}\check{\mathbf{t}}_j A_{jk}\right).
    \label{eq:forces}
\end{equation}
Cell stress is constructed from the first moments of $\mathsfbfit{M}$ which, from (\ref{eq:ALik}), are
\begin{subequations}
    \begin{align}
    {\textstyle{\sum_k}}\mathbf{r}_k \otimes \frac{\partial A_i}{\partial \mathbf{r}_k}&={\textstyle{\sum_{j,k}} }\tfrac{1}{2} \overline{A}_{jk}\mathbf{r}_k\mathbf{n}_{ij}^T= {\textstyle{\sum_{j}}} \mathbf{c}_j\otimes \mathbf{n}_{ij}=\mathsf{I}_2 A_i, \\
    {\textstyle{\sum_k}}\mathbf{r}_k \otimes \frac{\partial L_i}{\partial \mathbf{r}_k}&={\textstyle{\sum_{j,k}} }\overline{B}_{ij}A_{jk}\mathbf{r}_k\check{\mathbf{t}}_{j}^T= {\textstyle{\sum_{j}}}\overline{B}_{ij} \mathbf{t}_j\otimes \check{\mathbf{t}}_{j}\equiv \mathsfbfit{Q}_i L_i.
\end{align}
\label{eq:firstmoment}
\end{subequations}
The shape tensor $\mathsfbfit{Q}_i$ is defined in (\ref{eq:firstmoment}b) and satisfies $\mathrm{Tr}(\mathsf{Q}_i)=1$.

To build $\mathsfbfit{M}'$ we differentiate (\ref{eq:ALik}), giving
\begin{align}
    \frac{\partial^2 A_i}{\partial \mathbf{r}_k \partial \mathbf{r}_m}&=-\tfrac{1}{2}{\textstyle{\sum_j}} B_{ij} \overline{A}_{jk} A_{jm}\boldsymbol{\epsilon}_i, & 
    \frac{\partial^2 L_i}{\partial \mathbf{r}_k \mathbf{r}_m}&={\textstyle\sum_j}\overline{B}_{ij}A_{jk}A_{jm}\frac{\check{\mathbf{n}}_{ij}\otimes\check{\mathbf{n}}_{ij}}{t_j},
    \label{eq:mprime}
\end{align}
noting that $\check{\mathbf{n}}_{ij}\otimes \check{\mathbf{n}}_{ij}=\mathsf{I}-\check{\mathbf{t}}_{j}\otimes \check{\mathbf{t}}_{j}$ for cell $i$ neighbouring edge $j$.  It follows that $\partial^2 A_i/\partial \mathbf{r}_k^2=\mathbf{0}$ (because $\mathsf{B}\mathsf{A}=\mathsf{0}$).  For vertices $k$ neighbouring vertex $m$, with both neighbouring cell $i$ and $m\neq k$, $\partial^2 A_i/\partial \mathbf{r}_m\partial \mathbf{r}_k = \pm \tfrac{1}{2} \boldsymbol{\epsilon}_i$, with values decreasing in the direction of $\boldsymbol{\epsilon}_i$; for such vertices, $\partial^2 L_i/\partial\mathbf{r}_k \mathbf{r}_m$ takes the value $-\check{\mathbf{n}}_{ij}\otimes \check{\mathbf{n}}_{ij}/t_j$, where $j$ is the edge connecting $k$ and $m$.  

\section{Discrete spatial derivatives}
\label{app:hh}

  \begin{figure}
\begin{center}
    \includegraphics[width=\textwidth]{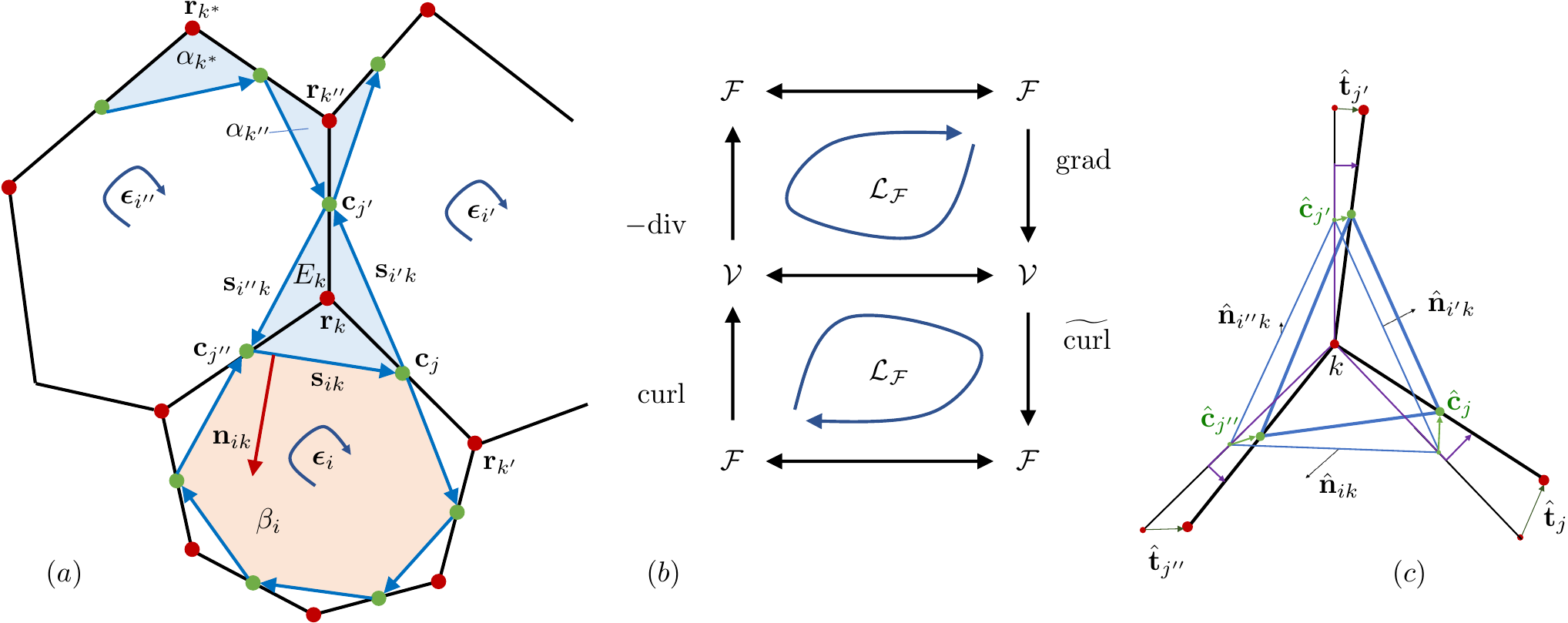}
\end{center}
    \caption{($a$) Cell vertices (red circles) are connected by edges (black lines); edge centroids (green circles) are connected by links (blue lines) defined by (\ref{eq:sik}) with normals $\mathbf{n}_{ik}$ defining $\partial A_i/\partial\mathbf{r}_k$ via (\ref{eq:Aik}).  Areas $E_k$ around vertices and $\beta_i$ within cells are shaded.  Two types of peripheral vertex are illustrated, labelled as $\mathbf{r}_{k''}$ and $\mathbf{r}_{k^*}$; $E_{k*}$ and $E_{k''}$ are shaded. The orientation $\boldsymbol{\epsilon}_i$ assigned to cell faces is here chosen to be clockwise in every cell.  ($b$) Operators on the edge-centroid network. $\mathrm{grad}$ is defined by (\ref{eq:grad}), its adjoint $-\mathrm{div}$ by (\ref{eq:div}) and the Laplacian over faces by (\ref{eq:lf}).  Curls are defined by (\ref{eq:curl}). Exact sequences (\ref{eq:exact}) are arranged vertically.  $(c)$ A small distortion of the network (from thin to thick edges) leads to changes $\hat{\mathbf{n}}_{ik}$ to the normals orthogonal to $\mathbf{s}_{ik}$, directed towards vertex $k$, along which pressures $\bar{P}_i$ act; changes to edges $\hat{\mathbf{t}}_j$ include rotation of unit vectors (purple arrows), along which tensions of adjacent cells $\bar{T}_{i}+\bar{T}_{i'}$ act.}
    \label{fig:hh}
\end{figure}

We consider how operators emerging from the vertex model relate to discrete approximations of conventional spatial derivaties over the network of links $\mathbf{s}_{ik}$ connecting adjacent edge centroids (see (\ref{eq:sik}) and Fig.~\ref{fig:hh}a).  
Following the scheme illustrated in Fig.~\ref{fig:hh}(b), we build the operators and their adjoints in terms of suitable inner products on $\mathcal{V}\subset \mathbb{R}^{N_v}\times \mathbb{R}^2$ (the vector space of vector-valued fields defined on vertices, equivalent to $\mathcal{TZ}$ in Fig.~\ref{fig:svdsummary}) and $\mathcal{F}\subset\mathbb{R}^{N_c}$ (the vector space of scalars defined on cell faces; this is distinct from $\mathcal{TY}$ in Fig.~\ref{fig:svdsummary} in holding a single attribute per cell), such that for all $\mathsfbfit{v}\in\mathcal{V}$, $\phi\in \mathcal{F}$ and $\psi\in\mathcal{F}$
\begin{equation}
    \langle\mathrm{grad}\,\phi,\mathsfbfit{v}\rangle_{\mathsf{E}}=\langle\phi,-\mathrm{div}\,\mathsfbfit{v}\rangle_{\mathcal{F}}, \quad
    \langle\mathrm{curl}\,\psi,\mathsfbfit{v}\rangle_{\mathsf{E}}=\langle\psi,\widetilde{\mathrm{curl}}\,\mathsfbfit{v}\rangle_{\mathcal{F}}.
    \label{eq:adj}
\end{equation}

$\langle \cdot,\cdot \rangle_{\mathsf{E}}$ is defined in (\ref{eq:inner0}), in terms of the area $E_k$ of the triangle at vertex $k$ bounded by links between adjacent edge centroids (Fig.~\ref{fig:hh}a).  $\langle\cdot;\cdot\rangle_{\mathcal{F}}$ is equivalent to the component of $\langle \cdot,\cdot \rangle_{\mathsf{G}^{-1}}$ acting on pressure variations, so that $\langle \mathsf{f},\mathsf{g}\rangle_{\mathcal{F}}={\textstyle{\sum_i}} A_i f_i g_i$.

At internal vertex $k$, the three adjacent $\mathbf{s}_{ik}$ form a closed triangle, with vertices at the edge centroids $\mathbf{c}_j$, circulating in the orientation congruent to $\boldsymbol{\epsilon}_k$ (Fig.~\ref{fig:hh}a).  For fixed $k$, the vectors $\mathbf{n}_{ik}=\boldsymbol{\epsilon}_i \mathbf{s}_{ik}$ form the three outward normals to the triangle surrounding vertex $k$.  For the three edge centroid links shown in Fig.~\ref{fig:hh}(a), ${\boldsymbol{\epsilon}_i \mathbf{s}_{ik}\cdot \mathbf{s}_{i'k}}=-2 E_k$ and ${\boldsymbol{\epsilon}_i \mathbf{s}_{ik}\cdot \mathbf{s}_{i''k}}=2 E_k$, which can be expressed more generally as
\begin{equation}
{\boldsymbol{\epsilon}_i \mathbf{s}_{ik}\cdot \mathbf{s}_{i'k}}=2 {E_k} {\textstyle \sum_j} A_{jk}B_{ij}\overline{B}_{i'j}. 
\label{eq:exid}
\end{equation}    
Applying the identity \cite{degoes2020} 

    $\int\nabla \phi \,\mathrm{d}A=\oint \phi \check{\mathbf{t}}\times \check{\mathbf{n}} \,\mathrm{d}s$

to the triangle, we can define for a field $\phi\in\mathcal{F}$ the discrete operator $\mathrm{grad}$ approximating $\nabla$ via
\begin{equation}
    \left\{ \mathrm{grad}\, \phi \right\}_k =\frac{1}{E_k} {\textstyle{\sum_i}} \boldsymbol{\epsilon}_i \mathbf{s}_{ik} \phi_i = \frac{1}{E_k}{\textstyle{\sum_{i,j}}}\tfrac{1}{2} \phi_i B_{ij} \boldsymbol{\epsilon}_i\mathbf{t}_j \overline{A}_{jk}=-\frac{1}{E_k}{\textstyle{\sum_i}} \frac{\partial A_i}{\partial \mathbf{r}_k}\phi_i,
\label{eq:grad}
\end{equation}
using (\ref{eq:Aik}).  This shows how $-\mathrm{grad}\,\mathsf{P}$ arises in the force balance (\ref{eq:forces}).  

From (\ref{eq:adj}), the adjoint operator to $\mathrm{grad}$ is, for all $\mathsfbfit{v}\in \mathcal{V}$,
\begin{equation}
    \left\{-\mathrm{div}\,\mathsfbfit{v}\right\}_i=\frac{1}{A_i} {\textstyle{\sum_k}}\boldsymbol{\epsilon}_i \mathbf{s}_{ik}\cdot\mathbf{v}_k=-\frac{1}{A_i}{\textstyle{\sum_k}} \frac{\partial A_i}{\partial \mathbf{r}_k}\cdot \mathbf{v}_k.
    \label{eq:div}
\end{equation}
We can verify that $\mathrm{div}$ is exact for linear functions of position, namely that 
\begin{equation}
\mathrm{div}\,(\mathsf{K}\cdot\mathbf{r}_k+\mathbf{J})=\mathrm{Tr}\,(\mathsf{K})
\label{eq:divexact}
\end{equation}
for a constant tensor $\mathsf{K}$ and vector $\mathbf{J}$, as follows.  $\sum_k -\boldsymbol{\epsilon}_i \mathbf{s}_{ik}\otimes \mathbf{r}_k$ can be viewed as a line integral around the polygon within cell $i$ bounded by links between edge centroids.  Since $\mathbf{r}_k$ is constant along each link, we can write the outward normal to each link as the sum of outward normals to half-edges adjacent to vertex $k$ using (\ref{eq:sik}), making the integral one around the periphery of cell $i$, so that 
\begin{equation}
-{\textstyle{\sum_k}} \boldsymbol{\epsilon}_i \mathbf{s}_{ik}\otimes \mathbf{r}_k =-{\textstyle{\sum_{j,k}}} \tfrac{1}{2}\boldsymbol{\epsilon}_i B_{ij}\mathbf{t}_j \overline{A}_{jk}\otimes \mathbf{r}_k=
-{\textstyle{\sum_{j}}} \tfrac{1}{2}\boldsymbol{\epsilon}_i B_{ij}\mathbf{t}_j \otimes \mathbf{c}_j
={\textstyle{\sum_{j}}} \mathbf{n}_{ij} \otimes \mathbf{c}_j = A_i \mathsf{I}
\label{eq:sr}
\end{equation}
where $\mathbf{n}_{ij}$ and $\mathbf{c}_j$ are as defined in Appendix~\ref{sec:appx}. This ensures that div satisfies (\ref{eq:divexact}), so that $\mathrm{div}\,\mathsfbfit{r}=2$ (which can be compared to (\ref{eq:mdiv})).  

We define complementary curls around vertices and faces as 
\begin{equation}
    \left\{\mathrm{curl}\,\mathsf{f} \right\}_k=\frac{1}{E_k}{\textstyle{\sum_i}} \mathbf{s}_{ik} f_i, \quad
    \left\{\widetilde{\mathrm{curl}}\,\mathsfbfit{v} \right\}_i=\frac{1}{A_i}{\textstyle{\sum_k}} \mathbf{s}_{ik}\cdot \mathbf{v}_k.
    \label{eq:curl}
\end{equation}
Multiplying (\ref{eq:sr}) by $\boldsymbol{\epsilon}_i$ and taking the trace confirms that $\widetilde{\mathrm{curl}}\,\mathsfbfit{r}=0$.  
To demonstrate that sequences in Fig.~\ref{fig:hh}(b) are exact, the identity (\ref{eq:exid}) 
implies that, when evaluating $-\mathrm{div}\,\circ \mathrm{curl}$ at cell $i$, the sum includes $-2f_{i'}$ from vertex $k$ and $2f_{i'}$ from vertex $k'$, leading to cancellation:
\begin{equation}
\left\{-\mathrm{div}\circ\mathrm{curl} \,\mathsf{f}\right\}_i=\frac{1}{A_i} {\textstyle\sum_{i', k}} \frac{\boldsymbol{\epsilon}_i \mathbf{s}_{ik}\cdot\mathbf{s}_{i'k} }{E_k} f_{i'}=
\frac{2}{A_i} {\textstyle\sum_{i',j, k}} A_{jk}B_{ij}\overline{B}_{i'j} f_{i'}=0
\end{equation}

because $\sum_k A_{jk}=0$.  Thus
\begin{equation}
    -\mathrm{div}\,\circ \mathrm{curl} =0, \quad \widetilde{\mathrm{curl}} \circ \mathrm{grad} =0.
    \label{eq:exact}
\end{equation}

We can therefore define a scalar Laplacian over faces as
\begin{equation}
    \left\{ \mathcal{L}_{\mathcal{F}} \mathsf{f} \right\}_{i} \equiv \left\{ -\mathrm{div}\,\circ \mathrm{grad} \, \mathsf{f} \right\}_i= \frac{1}{A_i} {\textstyle{\sum_{i',k}}} \frac{(\boldsymbol{\epsilon}_i \mathbf{s}_{ik})\cdot (\boldsymbol{\epsilon}_{i'}\mathbf{s}_{i'k})}{E_k} f_{i'}
= \frac{1}{A_i} {\textstyle{\sum_{i',k}}} \frac{ \mathbf{s}_{ik}\cdot \mathbf{s}_{i'k}}{E_k} f_{i'}.
    \label{eq:lf}
\end{equation}
We recognise $\mathcal{L}_\mathcal{F}$ as $\mathsf{A}_c^{-1}\mathcal{L}_A$; recall $\mathcal{L}_A$ is the first block of $\mathcal{L}$ in (\ref{eq:sprob}) and therefore $\mathcal{L}_\mathcal{F}$ is the first block of $\mathsf{G}\mathcal{L}\equiv \mathcal{L}_c^{\mathsf{G}\dag}$ in (\ref{eq:g}) and (\ref{eq:svo}).  $\mathcal{L}_{\mathcal{F}}$ has the form $ \mathsf{A}_c^{-1}\mathsf{B} \mathsf{F} \mathsf{B}^\top$ where $\mathsf{A}_c=\mathrm{diag}(A_1, \dots, A_{N_c})$ and $\mathsf{F}$ is diagonal, with elements defined along edges.  Its spectrum is illustrated in the inset to Fig.~\ref{fig:6}(b). Analogously, $\widetilde{\mathrm{curl}}\,\circ\mathrm{curl}=\mathcal{L}_\mathcal{F}$, as illustrated in Fig.~\ref{fig:hh}($a$), while the associated vector Laplacian mapping $\mathcal{V}\rightarrow \mathcal{V}$ is $\boldsymbol{\mathcal{L}}_{\mathcal{F}} \equiv
-\mathrm{grad}\,\circ \mathrm{div}+\mathrm{curl}\,\circ \widetilde{\mathrm{curl}}~$.  This system also admits a set of dual operators mapping scalars defined over vertices to vectors defined over faces.

Having identified the force due to pressure on a vertex with $\mathrm{grad}~\mathsf{P}$, where $\mathrm{grad}$ mimics $\nabla$, let us consider the corresponding force due to tension in (\ref{eq:forces}).  We define a new gradient operator mapping scalar fields on cells to vector fields on vertices, 
\begin{equation}
    \{\mathrm{grad}\,\mathsf{T}\}_k=-\frac{1}{E_k} {\textstyle\sum_{j}}  {\check{\mathbf{t}}_j} A_{jk} \overline{B}_{ij} T_i.
\end{equation}
This gives the vector sum of tensions acting along each edge radiating outward from vertex $k$.
Now defining $\langle f,g \rangle_\mathcal{F}\equiv \sum_j L_i f_i g_i$ (consistent with (\ref{eq:nonlinG})), then under inner products $\langle\cdot,\cdot\rangle_\mathsf{E}$ and $\langle \cdot,\cdot \rangle_{\mathcal{F}}$ we can define the adjoint operator as
\begin{equation}
    \{-\mathrm{div}\,\mathbf{v}\}_i= -\frac{1}{L_i}{\textstyle\sum_{j,k}}  \overline{B}_{ij} A_{jk}{\check{\mathbf{t}}_j}  \cdot\mathbf{v}_k.
\end{equation}
Summing over the vertices $k$ of cell $i$, this sums components of $\mathbf{v}_k$ pointing outwards from the cell.
The operator $-\mathrm{div}\,\circ\,\mathrm{grad}$ is a scalar Laplacian with element $ii'$ being

\begin{equation}
\frac{1}{L_i}{\textstyle{\sum_{j,j',k}}}\overline{B}_{ij} A_{jk} \frac{\check{\mathbf{t}}_j\cdot \check{\mathbf{t}}_{j'}}{E_k} A_{j'k} \overline{B}_{i'j'}.
\label{eq:lii}
\end{equation}
We recognise (\ref{eq:lii}) as $\mathsf{L}_c^{-1}\mathcal{L}_L$, where $\mathcal{L}_L$ is the final block of $\mathcal{L}$ in (\ref{eq:sprob}), making (\ref{eq:lii})) equivalent to the final block of $\mathsf{G}\mathcal{L}\equiv \mathcal{L}_c^{\mathsf{G}\dag}$ in (\ref{eq:g}) and (\ref{eq:svo}).  Fig.~\ref{fig:6}(b) shows how the spectrum of (\ref{eq:lii}) captures cell dilation modes when $\Gamma$ is large.  

\section{Gradient flow formulation}
\label{app:grfl}

Here we follow formalism presented by Peletier \cite{peletier2014}.  The state space $\mathcal{Z}=\mathbb{R}^{N_v}\times \mathbb{R}^2$ has elements $\mathsfbfit{r}(t)$ (in the sense that the evolution at any instant is determined entirely by vertex locations, for a given network topology).  Its tangent space $\mathcal{T}\mathcal{Z}$ at $\mathsfbfit{r}$ has elements $\mathsfbfit{v}(t)$, representing vertex velocities, so that $\dot{\mathsfbfit{r}}\in \mathcal{T}\mathcal{Z}$.  The energy $U$ is a map from $\mathcal{Z}$ to $\mathbb{R}$, when we consider ${U}=U(\mathsfbfit{r})$.  The (dual) cotangent space $\mathcal{T}\mathcal{W}$ has elements $\mathsfbfit{f}$ representing forces on vertices; we combine elements of the tangent and cotangent spaces via the scalar product (a bilinear form) $\mathsfbfit{f}^\top\cdot \mathsfbfit{v}\equiv \mathsfbfit{v}^\top \cdot \mathsfbfit{f}$.  The Fr\'echet derivative $U_\mathsfbfit{r}$ of the energy with respect to vertex displacements is defined via
\begin{equation}
    \lim_{s\rightarrow 0}\frac{U(\mathsfbfit{r}+s\mathsfbfit{v})-U(\mathsfbfit{r})}{s}=U_\mathsfbfit{r}^\top\cdot \mathsfbfit{v} \quad \mathrm{for~all}\quad \mathsfbfit{v}\in \mathcal{T}\mathcal{Z}.
\end{equation}
Thus $U_{\mathsfbfit{r}}(\mathsfbfit{r})\equiv \mathsfbfit{M}^\top \mathsf{g}\in \mathcal{T}\mathcal{W}$ and $\dot{U}\equiv (U_{\mathsfbfit{r}}(\mathsfbfit{r}))^\top \cdot \dot{\mathsfbfit{r}}=(\mathsfbfit{M}^\top \mathsf{g})^\top\cdot \dot{\mathsfbfit{r}}=\mathsf{g}^\top \mathsfbfit{M}\cdot\dot{\mathsfbfit{r}}$.

We write the evolution (\ref{eq:v1}) in canonical form as $\mathcal{D}(\mathsfbfit{r})\dot{\mathsfbfit{r}}=-U_{\mathsfbfit{r}}(\mathsfbfit{r})$. The dissipation operator $\mathcal{D}$ maps elements of $\mathcal{T}\mathcal{Z}$ to elements of $\mathcal{T}\mathcal{W}$.  Defining the bilinear form 
\begin{equation}
    (\mathsfbfit{v}_1,\mathsfbfit{v}_2)_{\mathcal{D},\mathsfbfit{r}}=\mathsfbfit{v}_1^\top \cdot \left( \mathsf{E}\otimes \mathsf{I}_2 + \mathsfbfit{M}^\top \mathsf{H}\mathsfbfit{M}\right)\cdot \mathsfbfit{v}_2,
    \label{eq:dr}
\end{equation}
we can write (\ref{eq:diss}) as $\dot{U}=-(\dot{\mathsfbfit{r}},\dot{\mathsfbfit{r}})_{\mathcal{D},\mathsfbfit{r}}\leq 0$.  Treating (\ref{eq:dr}) as an inner product (it is symmetric, bilinear and positive definite), we can write (\ref{eq:v1}) as $\dot{\mathsfbfit{r}}=-\mathrm{grad}_{\mathcal{D}}\, U\in \mathcal{T}_{\mathsfbfit{r}}\mathcal{Z}$.  In other words, 
\begin{equation}
(\mathsfbfit{v}, \dot{\mathsfbfit{r}})_{\mathcal{D},\mathsfbfit{r}}=-(\mathsfbfit{v}, \mathrm{grad}_\mathcal{D}\, U)_{\mathcal{D},\mathsfbfit{r}} \equiv  -\mathsfbfit{v}^\top\cdot U_{\mathsfbfit{r}}\quad \mathrm{for~all}\quad \mathsfbfit{v}\in \mathcal{T} \mathcal{Z}.
\end{equation}
This definition of $\mathrm{grad}_\mathcal{D}U$ shows how it is the Riesz representation of the Fr\'echet derivative $U_{\mathsfbfit{r}}$, and is determined by the dissipation inner product.

\section{Distinguishing cell shear and cell dilation modes}
\label{app:gam}

We write (\ref{eq:g}) in block form as
\begin{equation}
\left(\begin{matrix}
\mathsf{I}_{N_c}+\gamma_A \mathsf{A}_c^{-1} \mathcal{L}_A \mathsf{A}_c & \mathsf{A}_c^{-1} \mathcal{L}_C \mathsf{L}_{c}(\gamma_L/\Gamma L_0)\\ \mathsf{L}_{c}^{-1} \mathcal{L}_C^\top \mathsf{A}_c \gamma_A \Gamma L_0 & \mathsf{I}_{N_c}+\gamma_L \mathsf{L}_{c}^{-1} \mathcal{L}_L \mathsf{L}_{c} 
\end{matrix}
\right)\left(\begin{matrix}
 \dot{\mathsf{P}}  \\ \dot{\mathsf{T}}
\end{matrix}
\right)=-\left(\begin{matrix}
        \mathsf{A}_c^{-1}\mathcal{L}_{A} & \mathsf{A}_c^{-1}\mathcal{L}_C\\ \Gamma L_0 \mathsf{L}_{c}^{-1}\mathcal{L}_C^\top  & \Gamma L_0 \mathsf{L}_{c}^{-1}\mathcal{L}_L 
\end{matrix}
\right)\left(\begin{matrix}
   \mathsf{P} \\ \mathsf{T}
\end{matrix}
\right).
\label{eq:shd}
\end{equation}
For $\Gamma\sim \gamma_L\ll 1$, cell dilation modes are recovered by setting $\mathsf{T}= \Gamma \hat{T}$ in (\ref{eq:shd}) and discarding terms of $O(\Gamma)$, to give (\ref{eq:dilation}a) with $\mathsf{T}\approx \Gamma L_0 \mathsf{L}_{c}^{-1} \mathcal{L}_C^\top \mathcal{L}_A^{-1} \mathsf{A}_c \mathsf{P}$ (tensions are weak and are slaved to pressures).  Slower cell shear modes are recovered by setting $t=\tau/\Gamma$ in (\ref{eq:shd}), to give (at leading order in $\Gamma$) strong coupling between pressure and tension (but with opposite parity to cell dilation modes), $\mathsf{P}\approx -\mathcal{L}_A^{-1} \mathcal{L}_C T$, with $\dot{\mathsf{T}}\approx -\Gamma L_0 \mathsf{L}_{c}^{-1} (\mathcal{L}_L-\mathcal{L}_C^\top \mathcal{L}_A^{-1}\mathcal{L}_C)T$ (that is, the modes have the spectrum of the Schur complement $\mathsf{G}\mathcal{L}/(\mathsf{A}_c^{-1}\mathcal{L}_A)$).

For $\Gamma\sim \gamma_A^{-1}\gg 1$, cell dilation modes are recovered by setting $t=\tau/\Gamma$ and $P=\hat{P}/\Gamma$ in (\ref{eq:shd}), to give (\ref{eq:dilation}b) at leading order in $1/\Gamma$ with $\mathsf{P}\approx \mathsf{A}_c^{-1}\mathcal{L}_C \mathcal{L}_L^{-1} \mathsf{L}_{c} T/(\Gamma_0 L_0)$ (now, pressures are weak and are slaved to tensions).  Slower cell shear modes are recovered by taking dominant terms in (\ref{eq:shd}) for large $\Gamma$, giving $\mathsf{T}\approx -\mathcal{L}_L^{-1}\mathcal{L}_C^\top \mathsf{P}$ (strong coupling, with opposite parity to cell dilation modes) with $\dot{\mathsf{P}}\approx -\mathsf{A}_c^{-1}(\mathcal{L}_A-\mathcal{L}_C \mathcal{L}_L^{-1} \mathcal{L}_C^\top)P$ (the spectrum is that of the Schur complement $\mathsf{G}\mathcal{L}/(\Gamma L_0 \mathsf{L}_{c}^{-1}\mathcal{L}_L)$).

\section{Connecting prestress to stiffness}
\label{sec:prst}

Referring to Fig.~\ref{fig:hh}(c), (\ref{eq:lin}a) and (\ref{eq:mprime}), we can write the restoring force exerted on vertex $k$ by network prestress, under a small distortion of the network, as
\begin{multline}
-\left\{\bar{\mathsf{E}}^{-1}\bar{\mathsf{g}}^\top \bar{\mathsfbfit{M}'}\cdot\hat{\mathsfbfit{r}}\right\}_k=-\frac{1}{\bar{E}_k}\left[\bar{P}_i\hat{\mathbf{n}}_{ik}+\bar{P}_{i'}\hat{\mathbf{n}}_{i'k}+\bar{P}_{i''}\hat{\mathbf{n}}_{i''k}\right] \\
-\check{\mathbf{n}}_{ij} \frac{\check{\mathbf{n}}_{ij}\cdot \hat{\mathbf{t}}_{j}}{t_{j}} [\bar{T}_{i}+\bar{T}_{i'}]-\check{\mathbf{n}}_{i'j'} \frac{\check{\mathbf{n}}_{i'j'}\cdot \hat{\mathbf{t}}_{j'}}{t_{j'}} [\bar{T}_{i'}+\bar{T}_{i''}]-\check{\mathbf{n}}_{i''j''} \frac{\check{\mathbf{n}}_{i''j}\cdot \hat{\mathbf{t}}_{j''}}{t_{j''}} [\bar{T}_{i''}+\bar{T}_{i}].
\label{eq:prst}
\end{multline}
Recall $\mathbf{n}_{ik}$ is normal to $\mathbf{s}_{ik}$, the link in cell $i$ between adjacent edge centroids at vertex $k$, satisfying (\ref{eq:sik}).

Pressure of cell $i$ on vertex $k$ acts along $\mathbf{n}_{ik}=\boldsymbol{\epsilon}_i \mathbf{s}_{ik}$; distortion of the network changes the orientation of the normal as indicated by $\hat{\mathbf{n}}_{ik}$.  The distorted normals sum to zero ($\sum_i \hat{\mathbf{n}}_{ik}=0)$, indicating that the restoring force due to pressure vanishes if $\bar{P}_i=\bar{P}_{i'}=\bar{P}_{i''}$. An alternative expression for the pressure terms in (\ref{eq:prst}) is, using (\ref{eq:BAAid}),
\begin{equation}
    \frac{1}{E_k}\boldsymbol{\epsilon}_i\left[(\bar{P}_{i'}-\bar{P}_{i})\hat{\mathbf{c}}_j+(\bar{P}_{i''}-\bar{P}_{i'})\hat{\mathbf{c}}_{j'}+(\bar{P}_{i}-\bar{P}_{i''})\hat{\mathbf{c}}_{j''}\right]
\end{equation}
showing how pressure differences between cells act in the direction orthogonal to edge centroid displacements.  Tensions of the cells $i$ and $i'$ adjacent to edge $j$ act together along the unit vector pointing away from vertex $k$; distortion of the network rotates the unit vector, generating a force orthogonal to edge $j$.

\section{Stretch}
\label{app:stretch}

Consider a monolayer sitting on a membrane that undergoes a prescribed deformation.  We model adhesion between cells and the membrane by modifying the drag at vertices.  In a reference configuration, suppose the monolayer and membrane are in equilibrium with vertex locations $\mathbf{r}_k$.  These identify locations $\mathbf{R}_k$ on the membrane.  The membrane then undergoes biaxial stretch, mapping $\mathsfbfit{R}$ to $\boldsymbol{\Lambda}\cdot \mathsfbfit{R}$, where $\boldsymbol{\Lambda}\equiv \mathsf{I}_{N_v}\otimes (\Lambda_1(t)\mathbf{e}_1 \mathbf{e}_1^T +\Lambda_2(t) \mathbf{e}_2\mathbf{e}_2^T)$.  Here $\mathbf{e}_1$ and $\mathbf{e}_2$ are fixed orthogonal axes in the plane of the membrane and $\Lambda_1$ and $\Lambda_2$ are stretches.  The force balance (\ref{eq:v}) then becomes

$\mathsf{E} (\dot{\mathsfbfit{r}}-\dot{\boldsymbol{\Lambda}} \cdot \mathsfbfit{R})=-\mathsfbfit{M}^\top \mathsf{g}$,

penalising increasing displacements between $\mathsfbfit{r}$ and $\boldsymbol{\Lambda}\cdot \mathsfbfit{R}$,
dragging each vertex towards its counterpart on the membrane whenever there is membrane motion.  In this model, rapid membrane stretching promotes strong deformation of the monolayer.  Assuming that the membrane does not interfere with dissipation associated with area and perimeter changes, this leads to a forcing term $\mathsf{E}\dot{\boldsymbol{\Lambda}}\cdot \mathsfbfit{R}$ on the right-hand side of (\ref{eq:v1}).  Areas and perimeters then evolve according to a modified form of (\ref{eq:sprob}),
\begin{equation}
\left(\mathsf{I}_{2N_c}+\mathcal{L} \mathsf{H}\right)\dot{\mathsf{s}}=-\mathcal{L} \mathsf{g}+\mathsfbfit{M}\cdot \dot{\boldsymbol{\Lambda}}\cdot \mathsfbfit{R}, \quad \mathsf{g}=\mathcal{G}(\mathsf{s};\mathsf{s}_0,\Gamma).
\label{eq:sf}
\end{equation}
In Fig.~\ref{fig:stretch} we impose $\dot{\Lambda_1}=\pm \dot{\Lambda}_2=10^{-2}$ for $0<t<\tau$, and zero thereafter, for uniaxial ($-$) and biaxial ($+$) stretch.

%\end{appendix}

\bibliographystyle{RS}
\bibliography{abbreviated}

\end{document}